\documentclass[a4paper]{JAC2003}
\title{LONG-RANGE BEAM--BEAM EXPERIMENTS IN THE RELATIVISTIC HEAVY ION COLLIDER\\({\textnormal{Published in Phys. Rev. ST Accel. Beams 14, 091001}})}
\author{R.~Calaga, CERN, Geneva, Switzerland\\
W.~Fischer, N.~Milas, G.~Robert-Demolaize, BNL, Upton, NY, USA}
\usepackage[dvips]{epsfig,rotating}
\usepackage{multirow}
\usepackage{amsmath}
\usepackage{amstext}
\usepackage{amssymb}
\usepackage{color}
\usepackage{cite}

\usepackage{varwidth}
\usepackage{xcolor}

\begin{document}
\maketitle

\begin{abstract}
Long-range beam--beam effects are a potential limit to the LHC performance with
the nominal design parameters, and certain upgrade scenarios under discussion.
To mitigate long-range effects, current carrying wires parallel to the beam
were proposed and space is reserved in the LHC for such wires. Two current
carrying wires were installed in RHIC to study the effect of strong long-range
beam--beam effects in a collider, as well as test the compensation of a single
long-range interaction. The experimental data were used to benchmark
simulations. We summarize this work.
\end{abstract}

\section{Introduction}
The reader should note that this is an identical copy of an article
first published in~\cite{CALAGABBLR}.
Beam--beam effects have limited the performance of previous and existing
hadron colliders~\cite{Keil,Alex,JPK} such as the ISR~\cite{ISR0,ISR}, 
Sp$\mathrm{\bar{p}}$S~\cite{SPS0,
SPS1,SPS2,SPS3}, Tevatron~\cite{Zhan0,Shil3,Vali} and RHIC~\cite{RHIC1,RHIC2},
and are also expected to limit the performance of the LHC~\cite{Herr1,Gelf,
Gare,Furm,Grot,Leun,Zorz0,Zorz1,Herr0,LHC1,Zimm1,Kout2,Ster,Ohmi1,Dord4,Piel}.

Beam--beam effects can be categorized as either incoherent (dynamic aperture
and beam lifetime), PACMAN (bunch-to-bunch variations), or coherent (beam
oscillations and instabilities)~\cite{LHC1}. These effects can be caused
by both head-on and long-range interactions. Head-on effects, leading to
tune shifts and spreads, are important in all hadron colliders. Total
beam--beam induced tune shifts as large as 0.028 were achieved in the
Sp$\mathrm{\bar{p}}$S~\cite{SPS3} and Tevatron~\cite{Vali}, although
operational tune shift values are somewhat lower. The LHC in its early
stages of commissioning has already reached a total head-on beam--beam
tune shift of 0.02~\cite{WHEVIAN}.

Long-range effects, however, differ in previous and existing colliders.
In the ISR the beams collided under a large crossing angle of
15~deg~\cite{ISR1} that greatly reduced long-range effects.
In the Sp$\mathrm{\bar{p}}$S, with both beams in the same aperture and only
three bunches per beam, there were a few long-range interactions distributed over
the ring circumference. Due to the difference in the bunch intensities, the
effect on the anti-protons was stronger. In the Tevatron, also with both beams
in the same aperture but 36 bunches per beam, there are more long-range
interactions. With increased intensity of the anti-proton bunches, protons can
also be affected.

\begin{figure}[t]
\centering
\includegraphics[width=85mm]{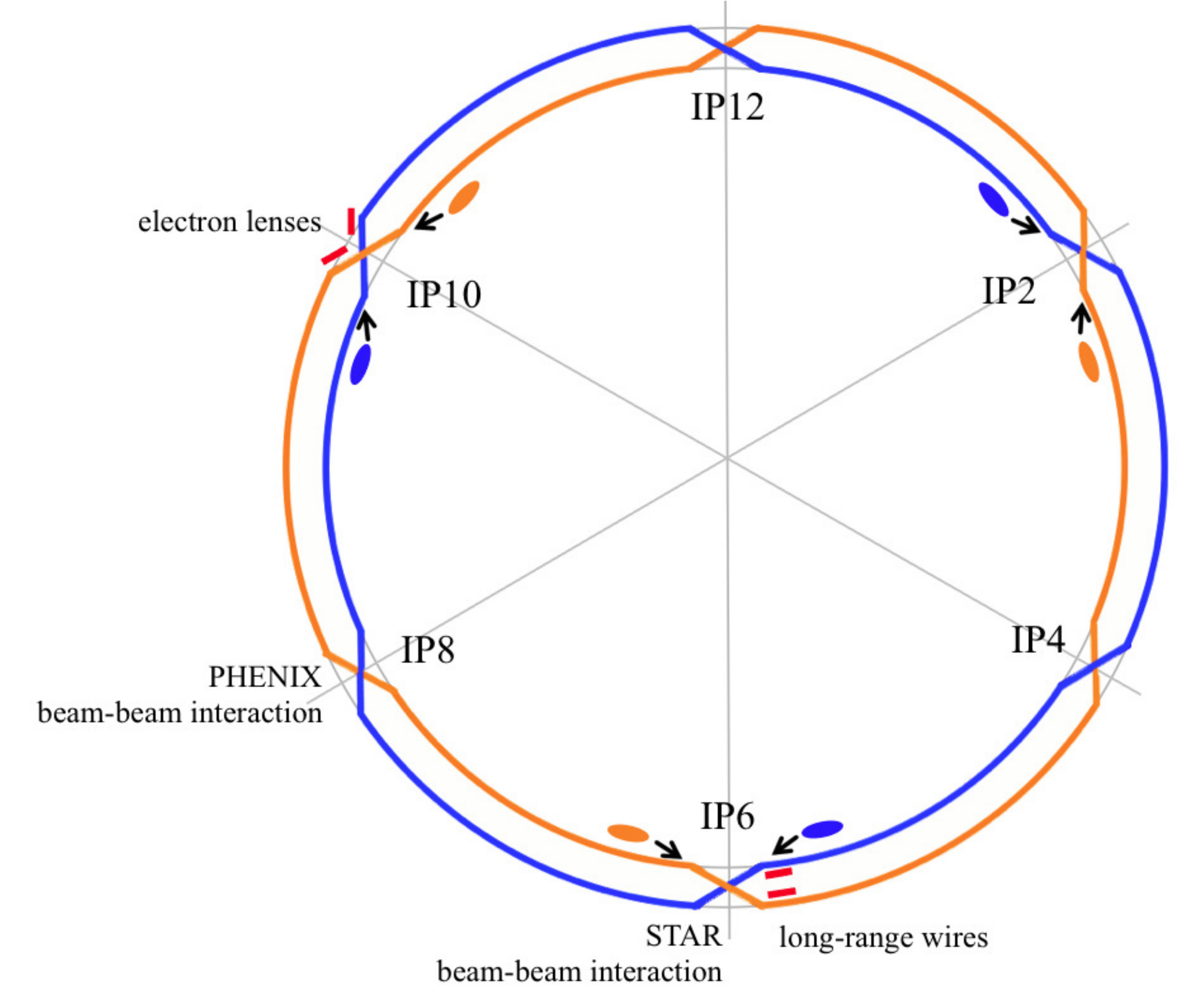}
\caption{Beam--beam interactions in RHIC and locations of wires and electron
lenses. }
\label{fig:01}
\end{figure}

In RHIC (Fig.~\ref{fig:01}), where both beams share the same
aperture only in the interaction regions, there is only one long-range
interaction per interaction region without an experiment
(a total of four in the current configuration),
with a 10~mm separation (corresponding to 30 rms beam
sizes for protons at 250~GeV energy). Long-range interactions have affected
the RHIC ramp transmission in the past~\cite{RHIC1}.

\section{Long-range effects and compensation in the LHC}
In the LHC there are 32 long-range beam--beam interactions
localized in each of four interaction regions~\cite{LHC1}.
A major upgrade of the LHC interactions region is foreseen by the end of the
decade with the primary objective to increase the average luminosity of the
machine by about a factor of 5 to 10 above the design performance. Among the
various upgrade scenarios a crab crossing scheme (CC), an early beam separation
scheme (ES), and a large Piwinski angle (LPA) are considered. In the CC
scheme, crab cavities placed on either side of the interaction region impart a
transverse kick to effectively compensate the crossing angle. This scheme
allows for a large crossing angle that greatly reduces long-range beam--beam
effects. In the ES scheme~\cite{Ster,Kout2} the number of long-range
interactions is greatly reduced but four parasitic collisions at 4--5~$\sigma$
per IP remain. In the LPA scheme~\cite{Zimm1} the small crossing angle will be
maintained, and long bunches of intensities up to $4$--$5\times 10^{11}$ protons
are used. All schemes aim at higher than nominal bunch currents and reduced
$\beta^*$. Therefore, long-range effects tend to become more problematic and
require more aperture for larger crossing angles or compensation to mitigate
these effects. The LPA scheme would most benefit from long-range beam--beam
compensation.
The compensation of long-range effects in the Tevatron was proposed with
electron lenses~\cite{Shil4}, and in the LHC with wires~\cite{Kout1}.
Electron lenses were also considered for the LHC~\cite{Dord2}, and
the use of wires was also studied for the Tevatron~\cite{Sen0}.
Implementation of long-range beam--beam compensation in the Tevatron is
challenging because the effect is distributed over the whole ring.
In the LHC the effect is localized in the interaction regions.
A partial long-range beam--beam compensation
was successfully implemented in the $\mathrm{e}^+\mathrm{e}^-$ collider
DA$\Phi$NE~\cite{Mila}. Beam--beam compensation and related issues were
reviewed at a workshop in 2007~\cite{Fisc3}.

\section{RHIC as a test bench for long-range studies}
\begin{figure}[h]
\centering
\fbox{\includegraphics[width=82mm]{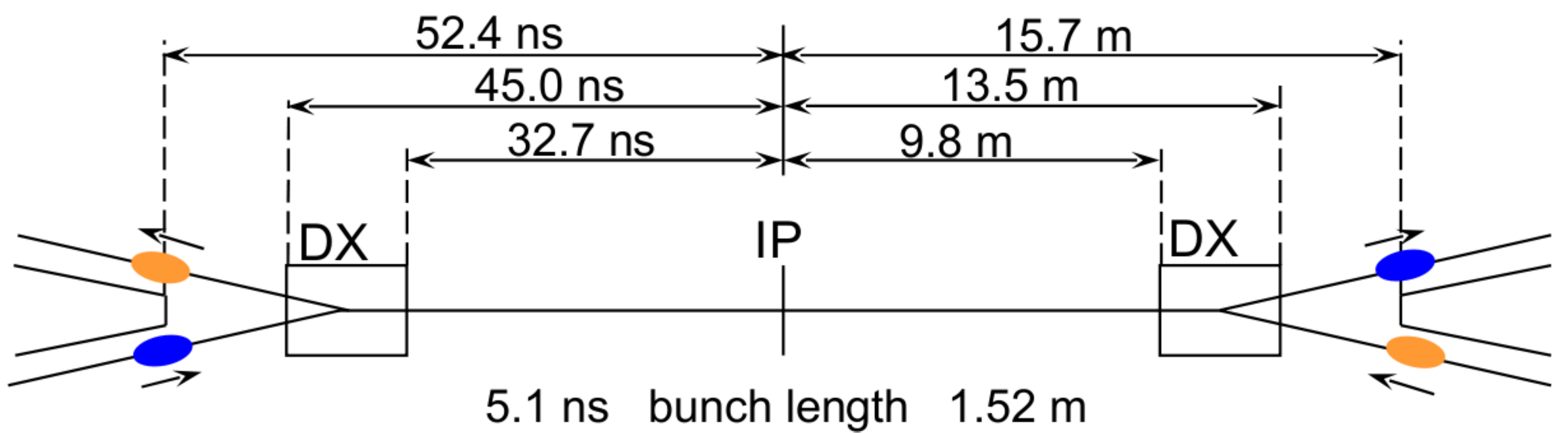}}
\caption{Schematic of the RHIC interaction regions.}
\label{fig:02}
\end{figure}

Figures~\ref{fig:01} and \ref{fig:02} show the basic layout of the beam--beam
interaction and compensation studies in RHIC. At store there are nominally
two head-on interactions in points 6 and 8 (IP6 and IP8),
and long-range interactions with a large separation in the other
interaction points. Three bunches in the Blue ring are coupled
to three bunches in the Yellow ring through the head-on beam--beam
interaction. For studies, two DC wires were installed in
the Blue and Yellow rings respectively in interaction region 6 (IR6).
Table~\ref{tab:RhicBB} shows the main beam parameters
for polarized proton operation, both achieved and design. In RHIC the
beam--beam effect is strongest in proton operation.
\begin{table}[htb]
\centering
\caption{Main RHIC parameters achieved in polarized proton operation that
are relevant for beam--beam effects protons (2009). Note that the polarized
proton bunch intensity is also limited by intensity dependent depolarization
effects in the AGS.}
\label{tab:RhicBB}
\small
\begin{tabular}{|l|c|c|c|}
\hline
Quantity                     & Unit       &          &        \\ \hline
Beam energy, $E_b$           & GeV        & 100      & 250    \\
Bunch intensity, $N_b$       & $10^{11}$  & 1.35     & 1.1    \\
Norm emittance, $\epsilon$   & $\mu$m     & 2.5      & 3.0    \\
rms bunch length, $\sigma_z$ & m          & 0.85     & 0.60   \\
Beam--beam parameter $\xi$/IP & ...        & 0.0056   & 0.0045 \\
No of IPs                    & ...        & 2        & 2      \\
$\beta^*$ at IP6, IP8        & m          & 0.7      & 0.7    \\
\hline
\end{tabular}
\normalsize
\end{table}

\begin{figure}[h]
\centering
\fbox{\includegraphics[width=70mm]{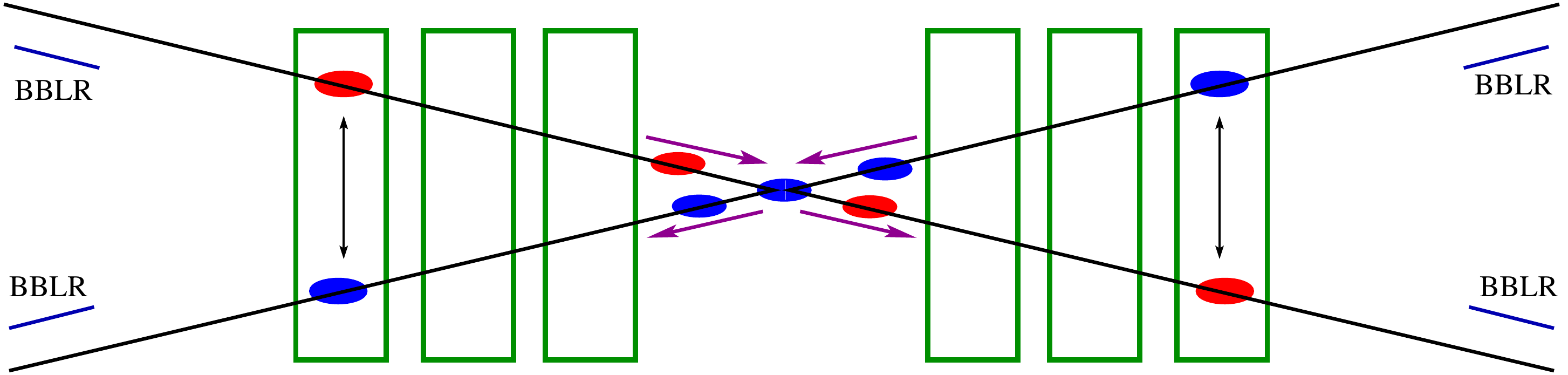}}
\includegraphics[width=48mm,angle=-90]{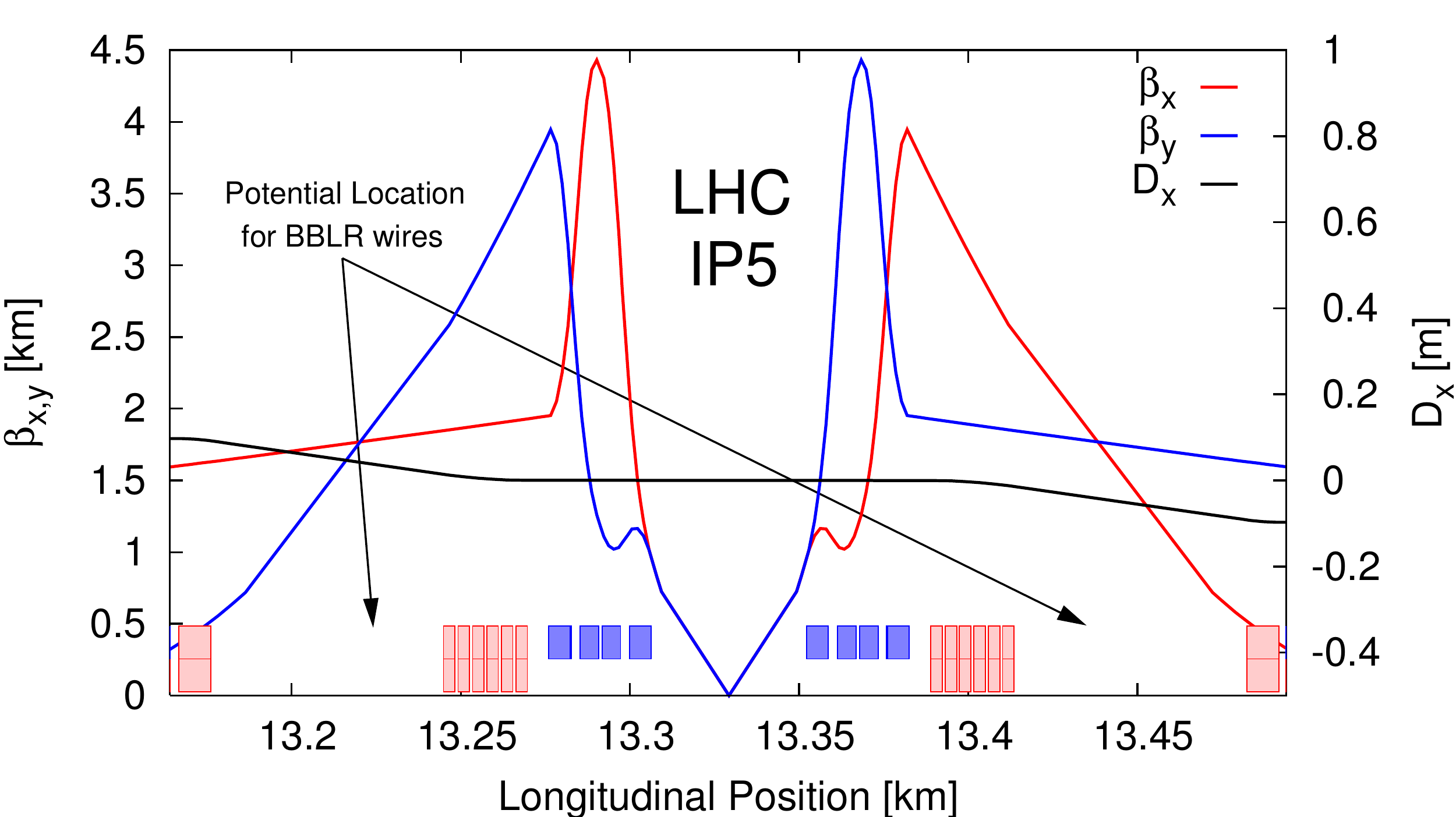}
\caption{LHC interaction region schematically showing the common
focusing channel with the 32 long-range interactions on the left and
the right of the collision point (top) and the optics functions in the
region.}
\label{fig:lhc}
\end{figure}

In the LHC locations in warm sections of the interaction regions
are reserved to
accommodate long-range beam--beam wire compensators (Fig.~\ref{fig:lhc}),
or electron lenses. These locations have about equal horizontal and vertical
$\beta$-functions.
With the expected strong long-range beam--beam effects in the LHC, and the
proposed wire compensation, experimental data and simulations of long-range
effects are highly desirable.
Operational and experimental data exist from the
Sp$\mathrm{\bar{p}}$S and the Tevatron. In the SPS, wires were installed to
further investigate strong long-range beam--beam interactions, to test the
compensation scheme, and to benchmark
simulations~\cite{Kout0,Zimm,Dord1,Dord4}.

The wire experiments in RHIC complement these studies. The beam lifetime
in RHIC is typical for a collider and better than in the SPS wire experiments.
In addition, and unlike in the SPS, head-on effects can be included, and with
properly placed long-range interactions and wires, the compensation of a
single long-range interaction is possible.

\section{Wires in RHIC}
The RHIC wire design is based on experience gained with the SPS units.
Design considerations are:
the location in ring, the integrated strength $(IL)$, the wire temperature
$T$ in operation, the positioning range and accuracy, power supply
requirements, controls, and diagnostics~\cite{Fisc0,Fisc1}. The wire
parameters are shown in Table~\ref{tab:bblr}.

\begin{table}[t]
\centering
\caption{Parameters for RHIC wires. The wire material is Cu at 20$^\circ$C.
The nominal wire strength is for a single long-range interaction with a proton
bunch intensity of $2\times 10^{11}$.}
\label{tab:bblr}
\small
\begin{tabular}{|l|c|c|}
\hline
Quantity                       & Unit   & Value \\
\hline
%{\bf input parameters}        &          \\
Strength $(IL)$, nominal       & A$\,$m & 9.6 \\
Max. strength $(IL)_{\mathrm{max}}$     & A$\,$m & 125 \\
Length of wire $L$             & m      & 2.5  \\
Radius of wire $r$             & mm     & 3.5  \\
Number of heat sinks $n$       & ...    & 3    \\
Electrical resistivity $\rho_e$& $\Omega\,$m & $1.72\!\!\times\!\! 10^{-8}$\\
Ceat conductivity $\lambda$    & W$\,$m$^{-1}$K$^{-1}$ & 384 \\
%density $\rho_g$               & kg/m$^3$  & $8.96\!\!\times\!\! 10^{3}$ \\
Thermal expansion coeff.       & K$^{-1}$  & $1.68\!\!\times\!\! 10^{-5}$\\
%melting temperature            & K         & 1083 \\
Radius of existing pipe $r_p$  & mm   & 60 \\ \hline
%{\bf output parameters}              &    &      \\
Current $I$, nominal                  & A  & 3.8  \\
Max. current $I_{\mathrm{max}}$                & A  & 50   \\
Current ripple $\Delta I/I$ (at 50 A) & $10^{-4}$ & $< 1.7$ \\
Electric resistance $R$               & m$\Omega$ & 1.12 \\
Max. voltage $U_{\mathrm{max}}$                & mV        & 55.9 \\
Max. power $P_{\mathrm{max}}$                  & W  & 2.8  \\
Max. temp. change $\Delta T_{\mathrm{max}}$    & K  & 15   \\
Max. length change $\Delta L_{\mathrm{max}}$   & mm   & 0.4  \\
Vertical position range               & mm/$\sigma_y$ & 65/10.6   \\
%vertical position range              & $\sigma_y$ & 10.6 \\
\hline
\end{tabular}
\normalsize
\end{table}

\subsection{Location in the Ring}~\label{sec:loc}
\begin{figure}[h]
\centering
\includegraphics[width=80mm]{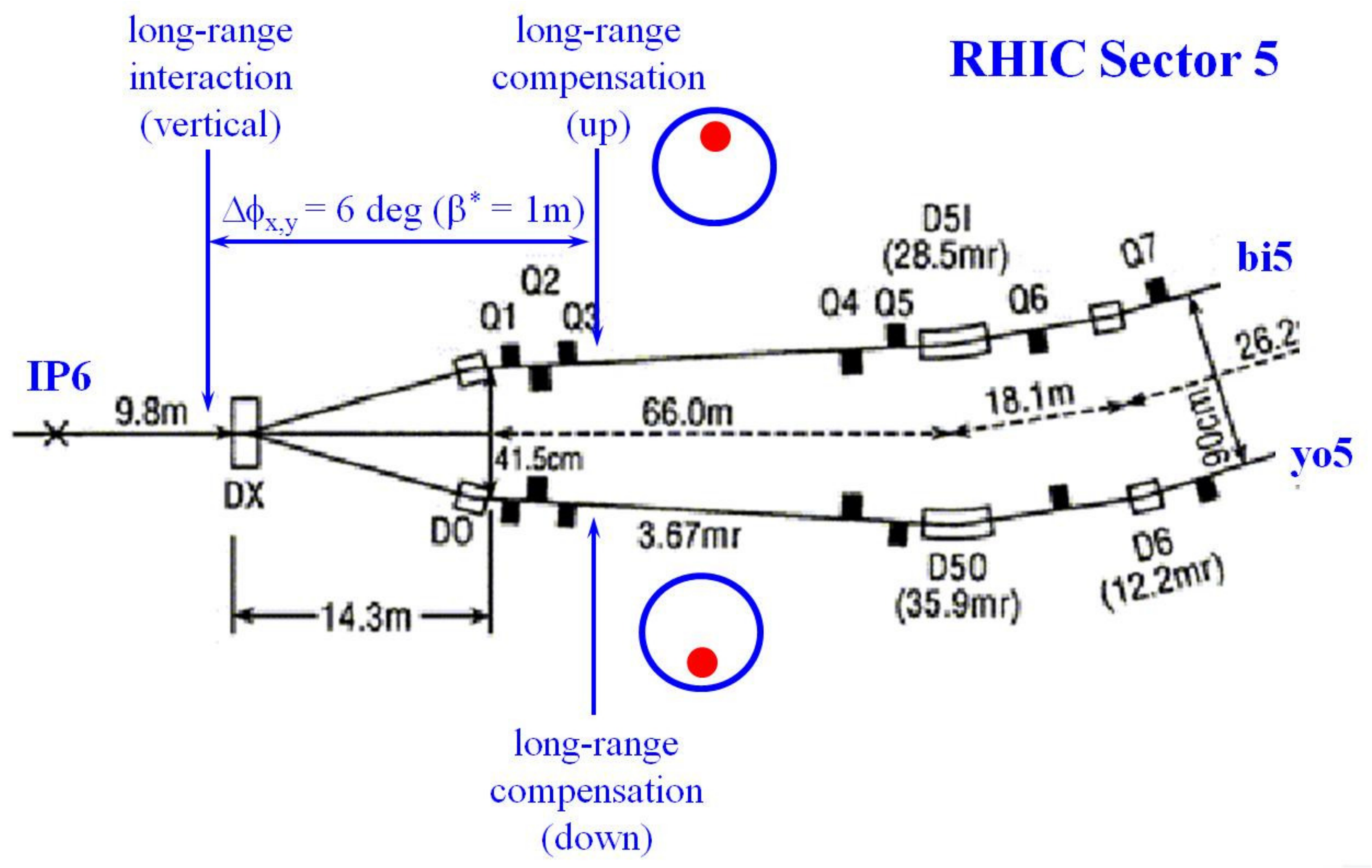}
\caption{Location of wires in RHIC and location of long-range beam--beam
interaction for compensation.}
\label{fig:ir6}
\end{figure}

\begin{figure}[h]
\centering
\includegraphics[width=80mm]{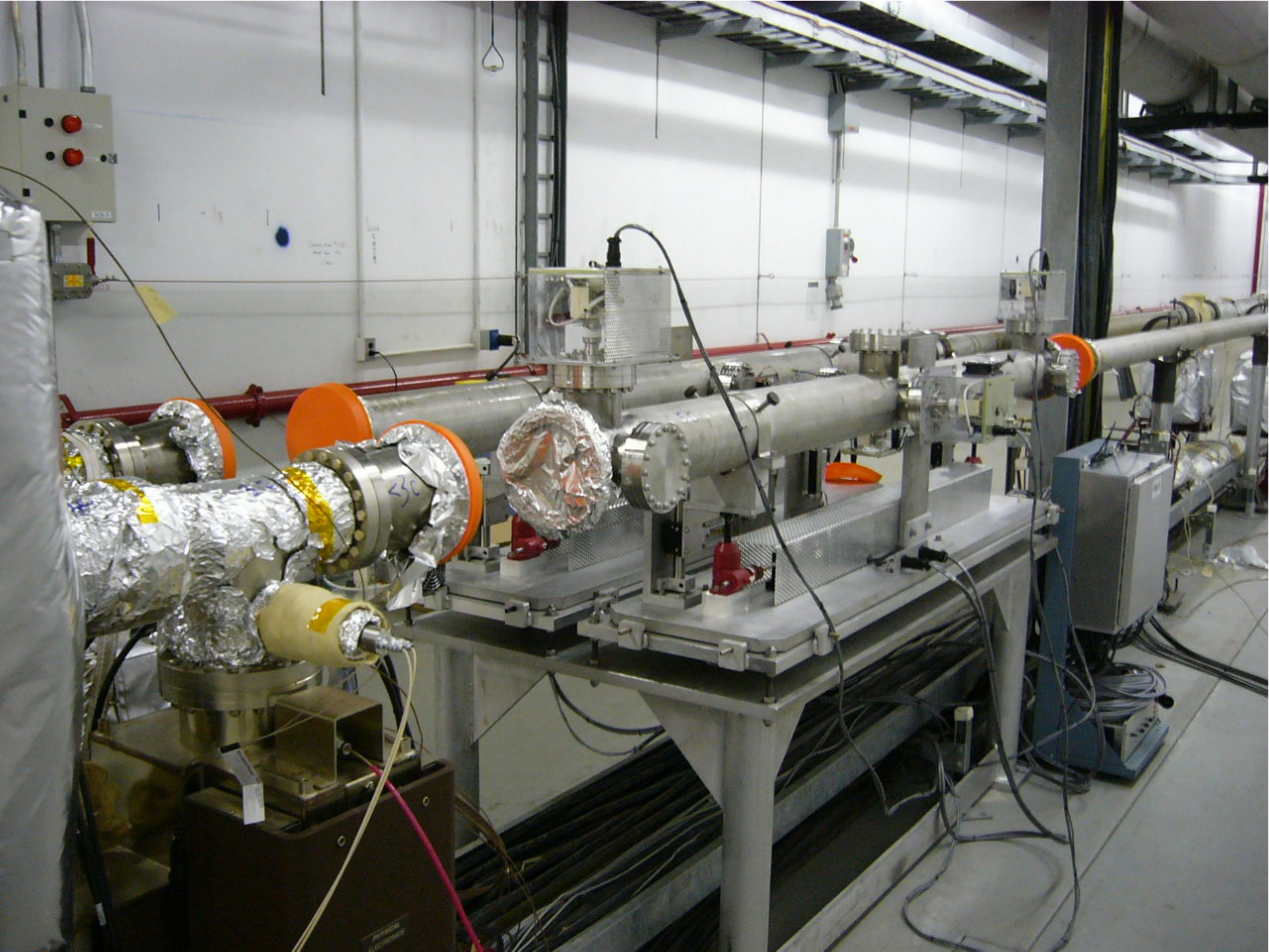}
\caption{The two long-range beam--beam wires in the RHIC tunnel during
installation.}
\label{fig:bblr_tunnel}
\end{figure}

For a successful compensation the phase advance between the long-range
interaction and the compensator should be no larger than about
10~degrees~\cite{Zimm9}. Lattices with $\beta^* \leq 1.0$~m have such small
phase advances between the entrance to the DX and the exit of Q3.
Thus it is possible to place a wire in the warm region after Q3 to
compensate for a long-range beam--beam interaction near the DX magnet
(Fig.~\ref{fig:ir6}). Since the beam paths must cross horizontally,
it is easier to control the distance between the beams in an experiment
through vertical separation. To compensate for a vertical long-range
interaction near the DX magnet, one wire can be installed in each ring
(see Fig.~\ref{fig:bblr_tunnel}). In the Blue ring the wire is installed
below the beam axis, in the Yellow ring above the beam axis.

\subsection{Integrated Strength}
To compensate a single long-range interaction, the
compensator's integrated strength $(IL)$ must be the same as the opposing
bunch's current integrated over its length $(IL) = N_bec$, where $I$ is the
current in the wire, $L$ its length, $N_b$ the bunch intensity, $e$ the
elementary charge, and $c$ the speed of light (see Table.~\ref{tab:bblr}).

In the LHC, an integrated strength of 80~A$\,$m is required to correct for the
16 long-range interactions on either side of an IR~\cite{Kout1}. Such a
strength is also expected to lead to enhanced diffusion at amplitudes larger
than 6~rms transverse beam sizes~\cite{Zimm9}. To study the enhanced diffusion
in RHIC, the wire is designed for $(IL)_{\mathrm{max}}=125$~A$\,$m.

\subsection{Wire Temperature}
The wire temperature should not exceed 100$^\circ$C to avoid increased
outgassing of the vacuum components. We use a number of air cooled heat
sinks to limit the wire temperature.

\begin{figure}[tbh]
\centering
\includegraphics[width=80mm]{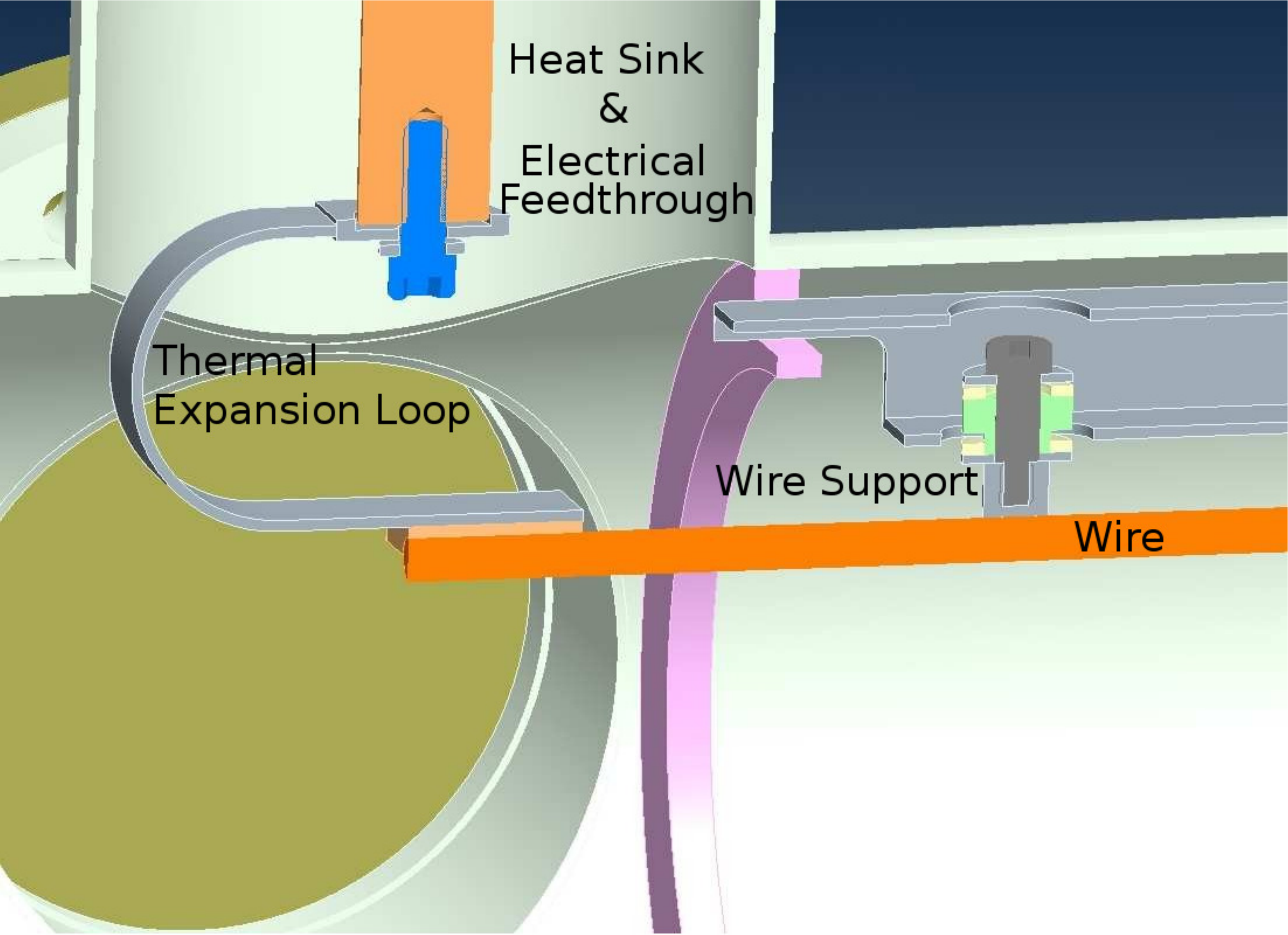}
\caption{Drawing of the end of a long-range beam--beam wire in RHIC.}
\label{fig:bblr_end}
\end{figure}

Assume first a wire in vacuum of radius $r$ and length $l$, with electrical
resistivity $\rho_e$ and heat conductivity $\lambda$. A current $I$ flows
through the wire, and at both ends there are heat sinks that maintain the
temperature $T_0$. Further we assume that the temperature rise $\Delta T$ in
the wire is small enough so that the material coefficients $\rho_e$ and
$\lambda$ are constants.
In each length element $\mathrm{d}x$ heat $\mathrm{d}Q$ is produced through the wire's
resistivity at the rate
\begin{equation}\label{eq:q1}
\frac{\mathrm{d}Q}{\mathrm{d}t} = \rho_e \frac{\mathrm{d}x}{\pi r^2} I^2,
\end{equation}
and the heat flow is connected to the temperature gradient $\mathrm{d}T(x)/\mathrm{d}x$ via
the heat equation
\begin{equation}\label{eq:q2}
 \frac{\mathrm{d}Q}{\mathrm{d}t} = -\lambda\pi r^2 \frac{\mathrm{d}T}{\mathrm{d}x}.
\end{equation}
Combining Eqs.~(\ref{eq:q1}) and (\ref{eq:q2}) yields the differential
equation for the temperature
\begin{equation}
  \frac{\mathrm{d}T^2(x)}{\mathrm{d}x^2} = - \frac{\rho_e}{\lambda}\frac{I^2}{\pi^2r^4}
\end{equation}
with the solution
\begin{equation}
  T(x) = -\frac{1}{2\pi^2}\frac{\rho_e}{\lambda}\frac{I^2}{r^4}x^2
  +ax+b.
\end{equation}
The coefficients $a$ and $b$ can be determined from the boundary conditions
$T(0)=T(l)=T_0$ yielding
\begin{equation}
 T(x) = T_0 + \frac{1}{2\pi^2}\frac{\rho_e}{\lambda}\frac{I^2}{r^4}
 (xl-x^2).
\end{equation}
The maximum temperature increase $\Delta T_{\mathrm{max}}$ is in the centre of the
wire, $x=l/2$, and is
\begin{equation}\label{eq:DTl}
 \Delta T_{\mathrm{max}} = \frac{1}{8\pi^2} \frac{\rho_e}{\lambda}
 \frac{(Il)^2}{r^4}.
\end{equation}
If we now assume a wire of length $L$ with $n$ heat sinks, we can replace
$l$ by $L/(n-1)$ in Eq.~(\ref{eq:DTl}) and arrive at
\begin{equation}
 \Delta T_{\mathrm{max}} = \frac{1}{8\pi^2} \frac{\rho_e}{\lambda}
 \frac{(IL)^2}{(n-1)^2 r^4}.
\end{equation}
We use $n=3$ heat sinks cooled with forced air. To move the wire compensator
close to the beam, its radius should not be much larger than an rms
transverse
beam size. The calculated temperature change is shown in Table~\ref{tab:bblr}.
Figure~\ref{fig:bblr_end} shows a drawing of the end of a wire.
Visible are the wire support, the electrical feed-through which is
also a heat sink, and a connecting loop allowing for thermal
expansion of the wire.

%\subsection{Wire temperature}
%The wire's temperature should not exceed 100$^\circ$C to avoid increased
%outgassing of the vacuum components. We use $n=3$ heat sinks cooled with
%forced air, spaced apart by $L/(n-1)$. The maximum temperature increase in
%the center between 2 heat sinks is
%\begin{equation}
%%\vspace*{-3mm}
% \Delta T_{max} = \frac{1}{8\pi^2} \frac{\rho_e}{\lambda}
% \frac{(IL)^2}{(n-1)^2 r^4},
%\end{equation}
%where $\rho_e$ is the electrical resistivity, $\lambda$ the heat conductivity,
%and $r$ the wire radius. To move the wire compensator close to the beam, its
%radius should not be much larger than an rms transverse beam size.
%The calculated temperature change with 3 heat sinks is shown in
%Tab.~\ref{tab:bblr}. Figure~\ref{fig:bblr_end} shows a drawing of the end
%of a wire. Visible are the wire support, the electrical feed-through which
%is also a heat sink, and a connecting loop allowing for thermal expansion of
%the wire.%

\subsection{Power Supply Requirements}
To limit emittance growth, a current ripple of $\Delta I/I<10^{-4}$ is
required~\cite{Zimm9}. A measurement shows a current ripple of
$\Delta I/I< 1.7\times10^{-4}$ where the upper limit is given by the
noise floor of the current measurement.

\section{Long-range experiments at RHIC}
More than 30 dedicated Long-Range (LR)  beam--beam experiments were performed at different
energies, with different species and various machine configurations.
They span a variety of long-range conditions which help
benchmark simulation tools.
The main parameters that were varied were the strength of the long-range
interactions (wire current), the distance between the beam and
the wire (or other beam), the tune and chromaticity.
All experimental sessions to study long-range beam--beam interactions
in RHIC can be broadly classified into three categories approximately in
chronological order:
\begin{itemize}
\item measurement of a single long-range interaction between the
  two proton bunches at 23 and 100 GeV in IP6;
\item effect of the DC wires on a single beam either by varying the current
  at a fixed distance or varying the distance to the beam with fixed current
  on both protons at 100 GeV and gold at 100 GeV/nucleon;
\item effect of long-range interaction either with a wire in the presence
  of head-on collisions or long-range interactions between the two
  beams in IP6 with simultaneous compensation using a wire at 100 GeV.
\end{itemize}

A summary of all long-range experiments performed in the RHIC accelerator
between 2005 to 2009 is listed with corresponding beam conditions
in Table~\ref{tab:LR}.
\begin{table*}[t]
\centering
\caption{Summary of long-range beam--beam experiments in RHIC. The wires in
the Blue and Yellow ring are named B-BBLR and Y-BBLR respectively. Fields
are left blank when the experimental value could not be determined.}
\label{tab:LR}
\scriptsize
\begin{tabular}{cccccccccccccl}
\hline\hline
fill   & \hspace*{-1mm} ring & \hspace*{-1mm} scan & \hspace*{-1mm} species & rel. & \hspace*{-4mm} bunches  &
$Q_x$  & $Q_y$  & LR       & LR       & LR         & fitted   & $d$ for  & comment \\
no     &        &          &          & $\gamma$   & \hspace*{-4mm} per ring &
       &        & location & strength & separation & exponent & $\tau<20$~h\\
       &        &          &          &            &          &
       &        &          & $(IL)$   & $d$        & $p$      &          \\
...    & ...    & ...      & ...      & ...        & ...      &
...    & ...    & ...      & A$\,$m   & $\sigma$   & ...      & $\sigma$ \\
\hline
\multicolumn{5}{l}{{\bf 2005}}\\
6981 & B & 1 & p  &  25.963 &  1 & 0.7331 & 0.7223 & IP4    &    5.3 & B moved      &      &      & weak signal \\
6981 & Y & 1 & p  &  25.963 &  1 & 0.7267 & 0.7234 & IP4    &    5.3 & B moved      &      &      & weak signal \\
6981 & B & 2 & p  &  25.963 &  1 & 0.7351 & 0.7223 & IP4    &    5.8 & B moved      &      &      & weak signal \\
6981 & Y & 2 & p  &  25.963 &  1 & 0.7282 & 0.7233 & IP4    &    5.8 & B moved      &      &      & weak signal \\
6981 & B & 3 & p  &  25.963 &  1 & 0.7383 & 0.7247 & IR4 DX &    8.6 & Y moved      &      &      & weak signal \\
6981 & Y & 3 & p  &  25.963 &  1 & 0.7271 & 0.7218 & IR4 DX &    8.6 & Y moved      &      &      & weak signal \\
6981 & B & 4 & p  &  25.963 &  1 & 0.7394 & 0.7271 & IR4 DX &    8.9 & Y moved      &  4.9 &  6.5 & \\
6981 & Y & 4 & p  &  25.963 &  1 & 0.7264 & 0.7388 & IR4 DX &    8.9 & Y moved      &  2.8 &      & \\
\hline
\multicolumn{5}{l}{{\bf 2006}}\\
7707 & B & 1 & p  & 106.597 & 10 &        &        & IR6 DX &    6.7 & B moved      &      &      & weak signal \\
7707 & Y & 1 & p  & 106.597 & 10 &        &        & IR6 DX &    6.7 & B moved      &      &      & weak signal \\
7707 & B & 2 & p  & 106.597 & 10 &        &        & IR6 DX &    6.7 & Y moved      &      &      & weak signal \\
7707 & Y & 2 & p  & 106.597 & 10 &        &        & IR6 DX &    6.7 & Y moved      &      &      & weak signal \\
7747 & B & 1 & p  & 106.597 &  8 &        &        & IR6 DX &    7.9 & B moved      &      &      & weak signal \\
7747 & Y & 1 & p  & 106.597 & 10 &        &        & IR6 DX &    7.9 & B moved      &      &      & weak signal \\
7747 & B & 2 & p  & 106.597 &  8 &        &        & IR6 DX &    7.0 & Y moved      &      &      & weak signal \\
7747 & Y & 2 & p  & 106.597 & 10 &        &        & IR6 DX &    7.0 & Y moved      &      &      & weak signal \\
7807 & B & 1 & p  & 106.597 & 12 & 0.6912 & 0.6966 & IR6 DX &    8.2 & Y moved      &  2.5 &  3.5 & additional octupoles\\
7807 & Y & 1 & p  & 106.597 & 12 & 0.7092 & 0.6966 & IR6 DX &    8.2 & Y moved      &  1.5 &  3.5 & additional octupoles\\
\hline
\multicolumn{5}{l}{{\bf 2007}}\\
8231 & B & 1 & Au &  10.520 &  6 & 0.2327 & 0.2141 & B-BBLR &   12.5 & B-BBLR moved &  7.2 &  6.5 & \\
8231 & B & 1 & Au &  10.520 &  6 & 0.2322 & 0.2140 & B-BBLR &    125 & B-BBLR moved &  7.8 &  9.0 & \\
8405 & B & 1 & Au & 107.369 & 56 & 0.2260 & 0.2270 & B-BBLR &    125 & B-BBLR moved &  1.7 & 15.0 & background test \\
8609 & B & 1 & Au & 107.369 & 23 & 0.2340 & 0.2260 & B-BBLR &   12.5 & B-BBLR moved &  7.4 &  6.0 & \\
8609 & B & 2 & Au & 107.369 & 23 & 0.2340 & 0.2260 & B-BBLR &    125 & B-BBLR moved & 16.0 &  5.5 & \\
8609 & Y & 1 & Au & 107.369 & 23 & 0.2280 & 0.2350 & Y-BBLR &   12.5 & Y-BBLR moved &  4.8 &  9.5 & \\
8609 & Y & 2 & Au & 107.369 & 23 & 0.2280 & 0.2350 & Y-BBLR &    125 & Y-BBLR moved &  4.1 &  7.5 & \\
8727 & B & 1 & Au & 107.369 & 23 & 0.2200 & 0.2320 & B-BBLR &   12.5 & B-BBLR moved &  5.2 &  9.5 & \\
8727 & B & 2 & Au & 107.369 & 23 & 0.2200 & 0.2320 & B-BBLR &    125 & B-BBLR moved &  8.1 & 10.0 & \\
8727 & B & 1 & Au & 107.369 & 23 & 0.2320 & 0.2280 & Y-BBLR &   12.5 & Y-BBLR moved &  6.3 &  4.5 & \\
8727 & B & 2 & Au & 107.369 & 23 & 0.2320 & 0.2280 & Y-BBLR &    125 & Y-BBLR moved & 10.8 &  5.0 & \\
8727 & B & 3 & Au & 107.369 & 23 & 0.2320 & 0.2280 & Y-BBLR &  125-0 & -6.5         &      &      & \\
8727 & B & 4 & Au & 107.369 & 23 & 0.2320 & 0.2280 & Y-BBLR &    125 & -6.5         &      &      & ver. chromaticity 2-8 \\
8727 & B & 5 & Au & 107.369 & 23 & 0.2320 & 0.2280 & Y-BBLR &  125-0 & -6.5         &      &      & ver. chromaticity 8 \\
\hline
\multicolumn{5}{l}{{\bf 2008}}\\
9664 & B & 1 & d  & 107.369 & 12 & 0.2288 & 0.2248 & B-BBLR &    125 & B-BBLR moved &  3.8 & 17.0 & end of physics store\\
9664 & B & 2 & d  & 107.369 & 12 & 0.2288 & 0.2248 & B-BBLR & 75-125 & 5.8          &      &      & end of physics store\\ \hline
\multicolumn{5}{l}{{\bf 2009}}\\
10793& B & - & p  & 106.597 & 36 & 0.691  & 0.688  & B-BBLR & 125    & B-BBLR moved &     &      & with head-on collisions \\
10793& Y & - & p  & 106.597 & 36 & 0.695  & 0.692  & Y-BBLR & 125    & Y-BBLR moved &     &      & with head-on collisions \\
10793& B & - & p  & 106.597 & 36 & 0.691  & 0.688  & IR6 DX & 12.5   & B-BBLR moved &     &      & LR compensation \\
10793& Y & - & p  & 106.597 & 36 & 0.695  & 0.692  & IR6 DX & 12.5   & Y-BBLR moved &     &      & LR compensation \\
\hline\hline
\end{tabular}
\normalsize
\end{table*}
The main observables in long-range beam--beam experiments are orbits,
tunes, Beam Transfer Functions (BTFs), and the beam lifetime.
Several simulations were performed for a subset of measurements which
show successful reconstruction of all measurable quantities and the onset
of losses~\cite{Fisc2}. Specific examples for each of the three categories
with detailed results are presented in the next sections
to summarize all the long-range experiments performed at RHIC.
%to summarize the The three categories of experiments with specific
%examples are detailed in the next sections
%with specific measurements and corresponding results.

\subsection{Single Long-range Measurements}
The first set of long-range beam--beam experiments were performed
with proton beams in 2006. The motivation of these experiments was to
characterize the effect of one parasitic interaction on beam losses
for a future compensation demonstration. The Blue and Yellow beams
were vertically separated in the IR6 region close to the DX magnet
(Fig.~\ref{fig:02}). The RHIC beams are very stable at the nominal
working point and the effect of a single long-range (weak effect)
is not visible in the beam lifetime. An effect of a
compensation effect will not be possible to detect with the available
instrumentation.

\begin{figure}[htb]
\centering
\includegraphics[width=5.9cm,angle=-90]{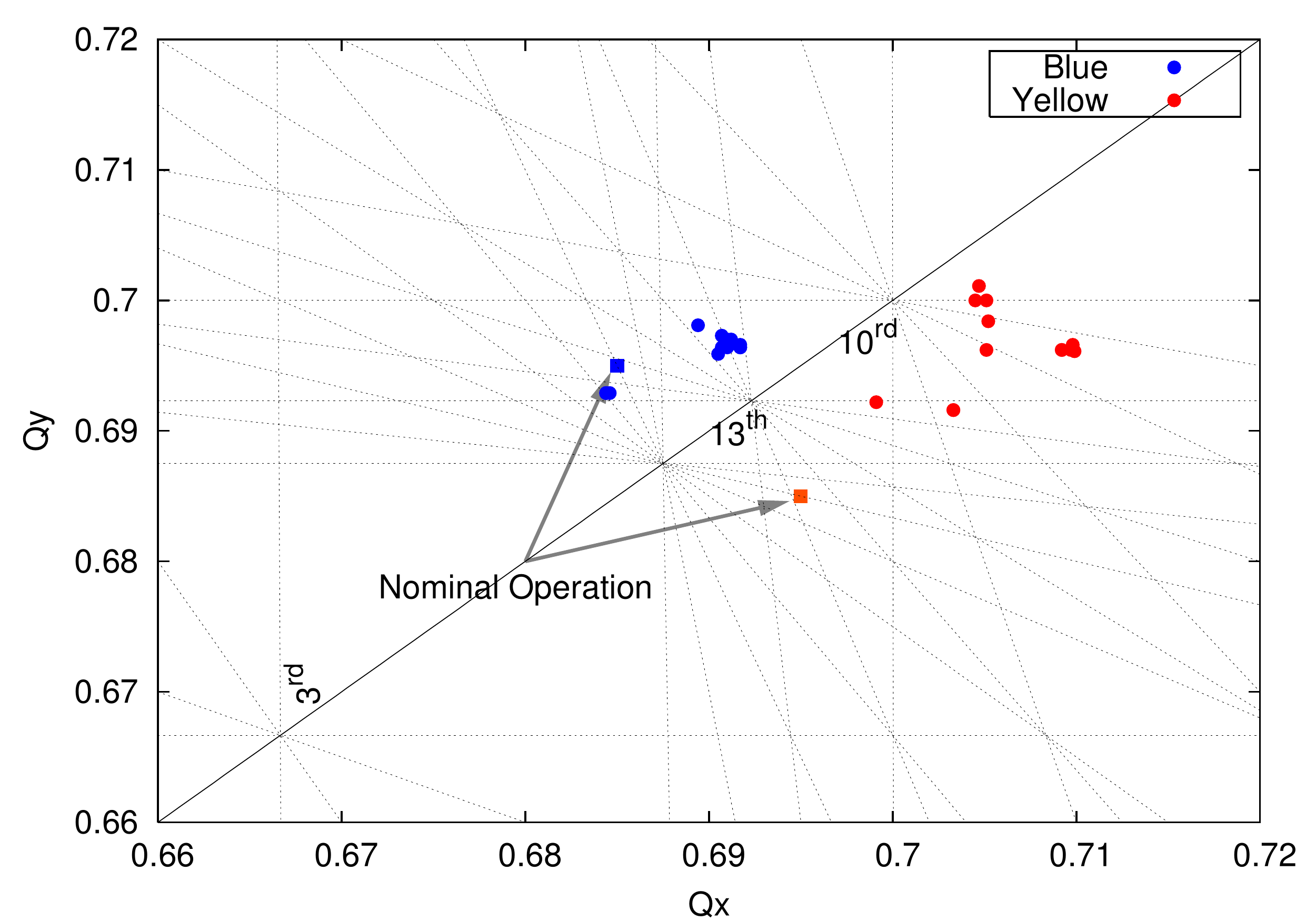}
\caption{Tunes in the resonance diagram for both beams and both planes
during a scan.}
\label{fig:tplot}
\end{figure}

\begin{figure}[htb]
\centering
\includegraphics[width=4.8cm,angle=-90]{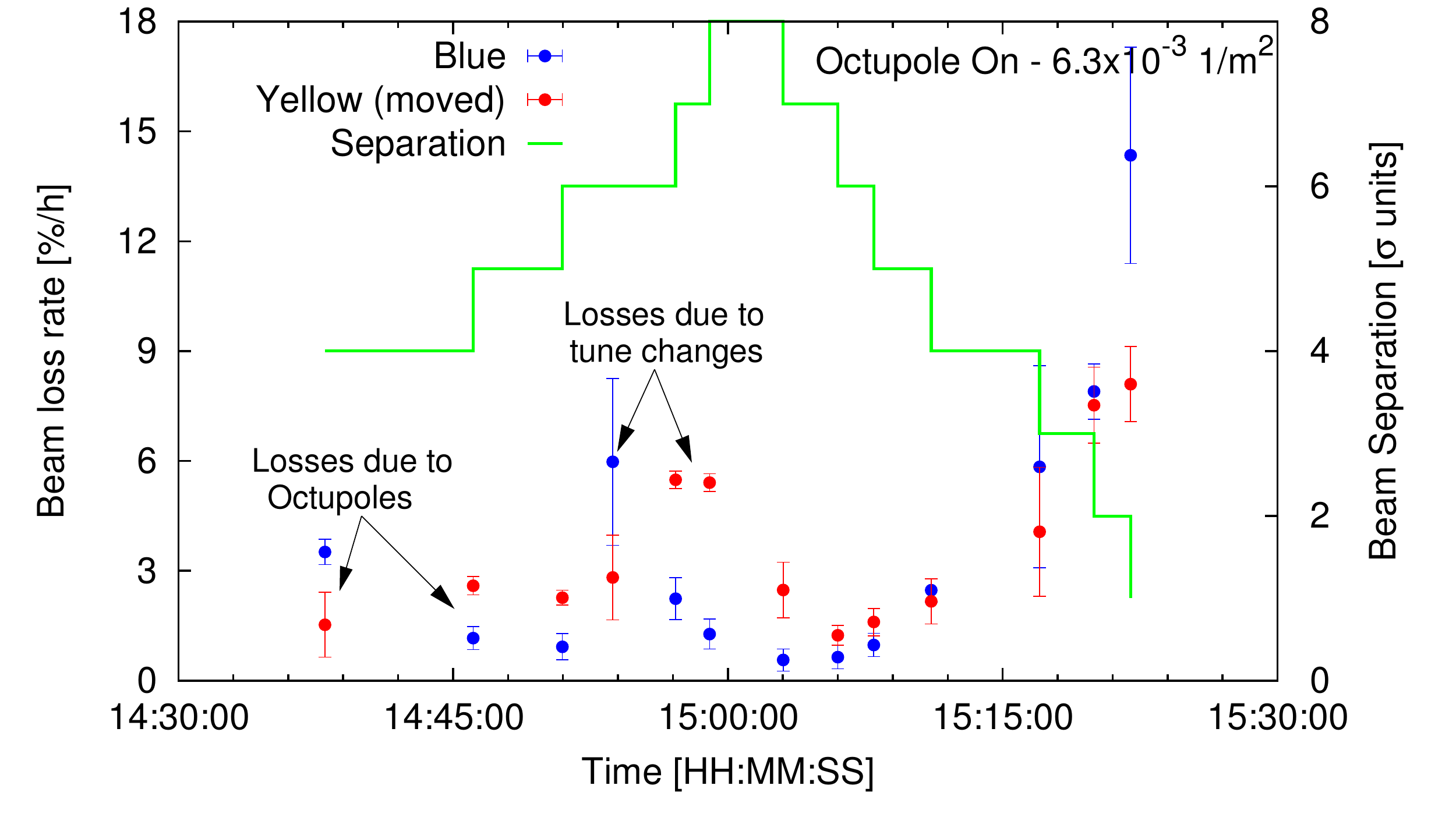}
\caption{Beam losses due to a single parasitic interaction
of the Blue and Yellow beam. The Yellow beam moved closer to the Blue
beam from an initially large separation.}
\label{fig:SLL}
\end{figure}

Therefore, a finite strength in lattice octupoles and a working point
close to the 10$^{\text{th}}$ order resonance was used as shown in
Fig.~\ref{fig:tplot}. At this modified working point,
the beams are marginally stable as the introduction
of the single parasitic interaction increases the tune spread
of the large amplitude particles on to the 10$^{\text{th}}$ order resonance,
thus enhancing the effect. This setup of marginally stable beams is only
used for experiments with single long-range interaction between the two
beams.
Some relevant lattice and beam parameters are
listed in Table~\ref{tab:bblr1}.
\begin{table}[htb]
\centering
\caption{RHIC parameters for experiments with long-range interactions
with proton beams.}
\label{tab:bblr1}
\small
\begin{tabular}{|l|c|c|c|}
\hline
Quantity  & Enit & Blue & Yellow \\
\hline
Beam energy $E$             & GeV/n  & \multicolumn{2}{c|}{100}  \\
Rigidity $(B\rho)$          & T$\,$m & \multicolumn{2}{c|}{831.8}\\
Number of bunches           & ...    & \multicolumn{2}{c|}{12}   \\
LR interaction from IP6     & m      & \multicolumn{2}{c|}{10.6} \\
Norm. Emittances ($\epsilon_{x,y}$)  & $\mu$m & \multicolumn{2}{c|}{15-20} \\
$\beta_x$ at wire location  & m      & \multicolumn{2}{c|}{105} \\  \cline{3-4}
Tunes (Q$_{x,y}$)           & ...    & 0.69/0.7   & 0.71/0.69 \\ \hline
$\beta_x$ at wire location  & m      & 1060  & 342 \\
$\beta_y$ at wire location  & m      & 357   & 1000 \\ \hline
Octupole Strength (kl)          & m$^{-2}$
& \multicolumn{2}{c|}{6.3 $\times 10^{-3}$} \\  \hline
\end{tabular}
\normalsize
\end{table}
The marginally stable beams were essential as the effect of the single
long-range interaction on the rather stable RHIC beams is subtle. In one such
experiment, the effect on the beam losses on both beams as a function of the
separation is shown in Fig.~\ref{fig:SLL}. To increase the signal-to-noise
ratio the losses are averaged over the 12 bunches.

Note that the Yellow beam was moved while the Blue beam was kept stationary.
Therefore, the effect on the Blue beam is of relevance as the losses in the
Yellow beam may also be affected by orbit and tunes shifts. A small effect is
visible when the beams are approximately 5~$\sigma$ or closer.

Compensation of such small effects is difficult as the losses are smaller than
the natural reproducibility of the machine for a given beam setup. Therefore,
it was important to significantly enhance the loss due to the long-range
interactions to clearly demonstrate compensation with a DC wire. Increased
chromaticity and introduction of head-on collisions were utilized to enhance
the effect of the LR interaction with the DC wires~\cite{Fisc2}.

\subsection{Wire Scans on Single Beam}
After the installation of the DC wires in 2007, the majority of the experiments
were carried out using the individual wires of the Blue and the Yellow ring to
characterize the onset of the losses under certain beam
conditions~\cite{Fisc1,Fisc2}. Most of the wire experiments were done with
gold beams. Table~\ref{tab:bblrAU} shows the main beam parameters for the wire
experiments at store with gold beams.

\begin{table}[htb]
\centering
\caption{RHIC parameters for experiments with DC wires on individual
gold beams.}
\label{tab:bblrAU}
\begin{tabular}{|l|c|c|c|}
\hline
Quantity  & Unit & Blue & Yellow \\ \hline
Beam energy $E$          & GeV/nucleon  & \multicolumn{2}{c|}{100}  \\
Rigidity $(B\rho)$             & Tm     & \multicolumn{2}{c|}{831.8}\\
Number of bunches              & ...    & \multicolumn{2}{c|}{6--56} \\ \cline{3-4}
Norm. Emittance $\epsilon_{x,y}$   & $\mu$rad & 17 &  17 \\ \hline
Distance IP6 to wire & m      & \multicolumn{2}{c|}{40.92} \\
centre    & & \multicolumn{2}{c|}{}\\
Parameter $K$ (at 50~A)        & nm     & \multicolumn{2}{c|}{$-30.1$} \\
Hor. tune Q$_x$                & ...    & 28.234    & 28.228 \\
Ver. tune Q$_y$                & ...    & 28.226    & 29.235 \\
$\beta_x$ at wire location     & m      & 1091  & 350 \\
$\beta_y$ at wire location     & m      & 378  & 1067 \\ %\hline
\hline
\end{tabular}
\normalsize
\end{table}

\begin{figure}[htb]
\centering
\includegraphics[width=58mm,angle=-90]{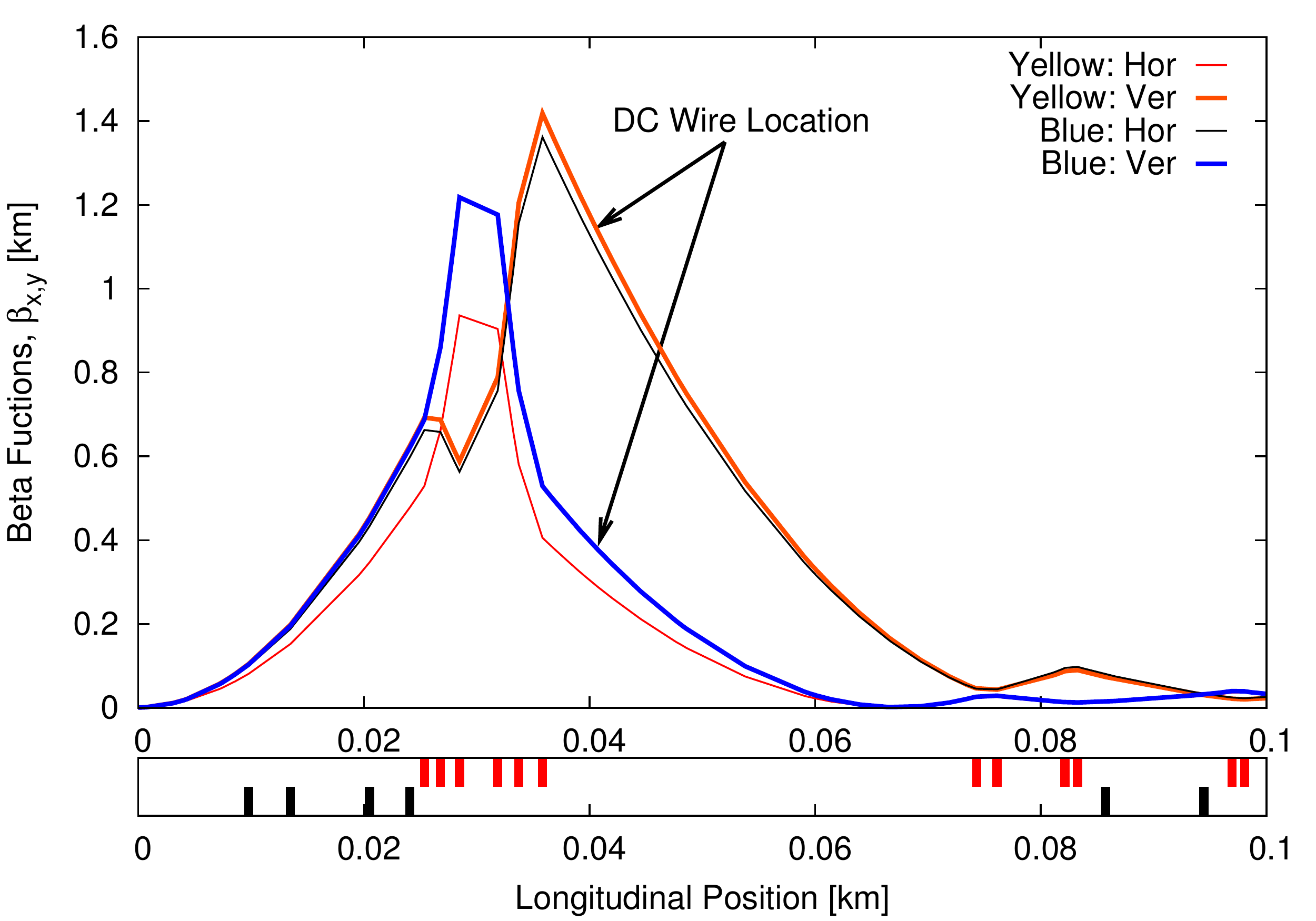}
\caption{The beta functions from MAD lattice for both rings as a function of
longitudinal position.}
\label{fig:OPTICS}
\end{figure}

The $\beta$-functions in Table~\ref{tab:bblrAU} are the best estimate of the
real $\beta$-functions in the machine. The design lattice has $\beta^*=0.8$~m
at IP6. To calculate the $\beta$-functions at the wire location
we use $\beta^*=0.9$~m, and assume a 10\% error. Figure~\ref{fig:OPTICS} shows
the MAD lattice near the interaction region 6 where the wires are located.

\begin{figure}[htb]\centering
\includegraphics[width=48mm,angle=-90]{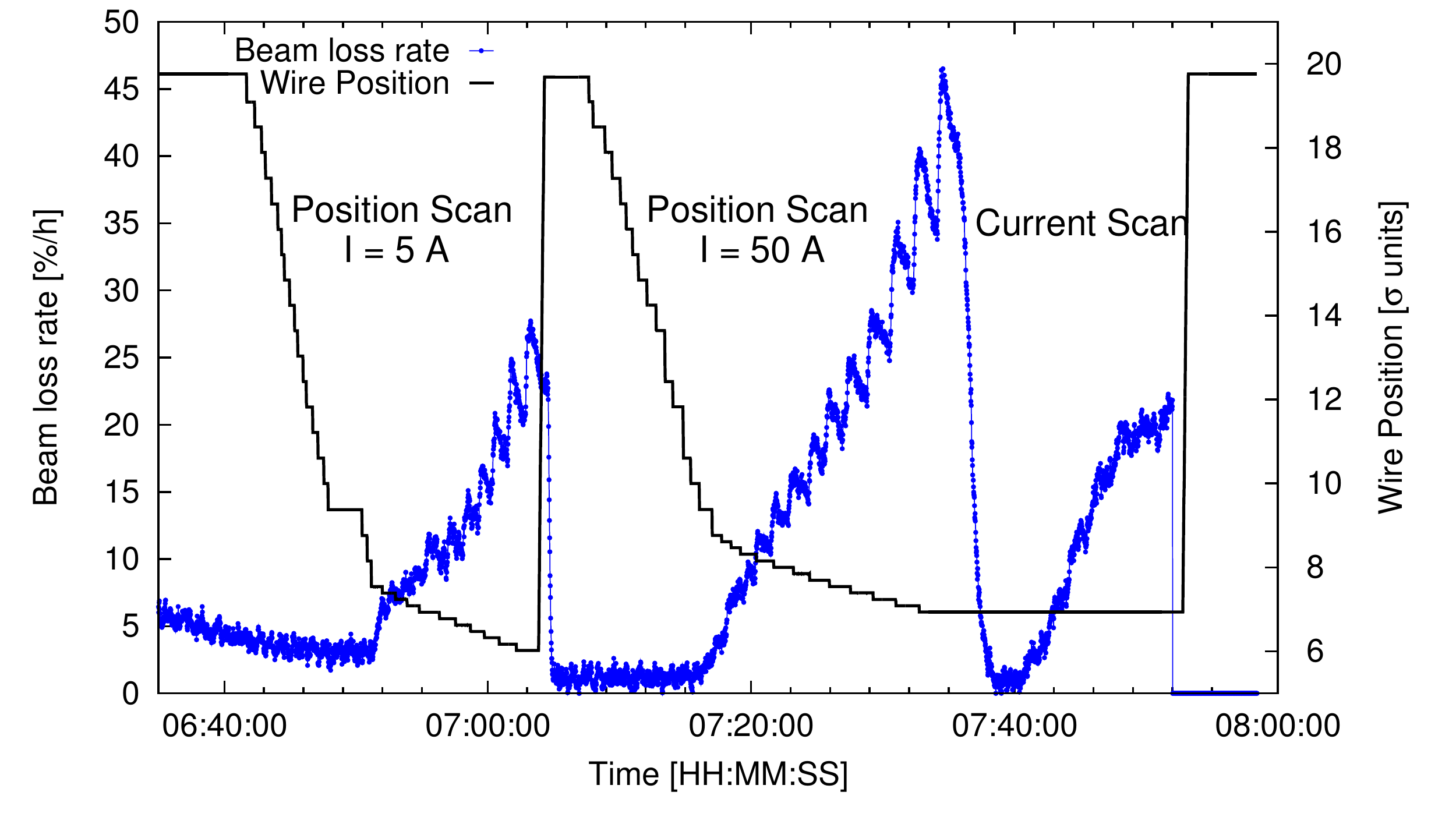}
\includegraphics[width=48mm,angle=-90]{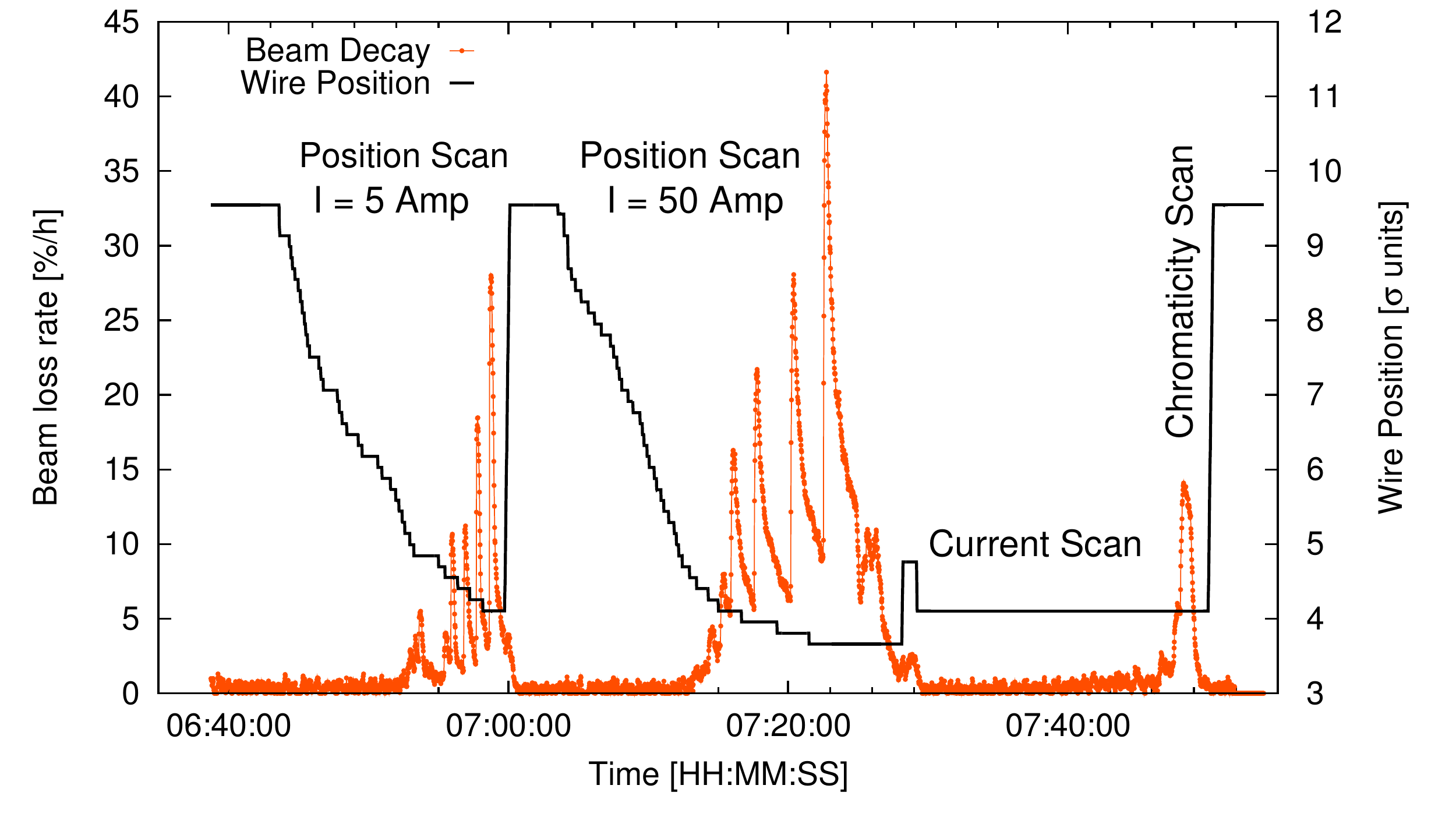}
\caption{Beam losses and wire position as a function of time
during the course of position and current scan (May 2007, Fill 8727).}
\label{fig:POSSCAN}
\end{figure}

The measurements consisted mainly of distance and current scans and
simultaneous measurements of the beam loss rate. An overview of the beam losses
and wire position for the Blue and the Yellow ring during the course of a scan
(Fill 8727) is illustrated in Fig.~\ref{fig:POSSCAN}.
The beam loss rates are clearly
different for the Blue and Yellow beams. This indicates towards different
diffusion rates and re-population of tails for the two beams. The exact reason
for this difference is not identified. It should be noted that the wire
installations are identical.

Orbit, tune and chromaticity changes can be calculated as a function of the
long-range strength and distance~\cite{Sen1}. These quantities and beam
transfer functions are usually recorded to benchmark with theory and
simulations. The vertical dipole kick $\Delta y'$ and vertical tune change
$\Delta Q_y$  due to the wire for a separation $d$ in the vertical plane
between the beam and the wire are given by (assume no horizontal separation)
\begin{equation}\label{eq:tune}
  \Delta y' = \frac{K}{d}
  \;\;\;\;\; \mathrm{and} \;\;\;\;\;
  \Delta Q_{x,y} = \pm\frac{K\beta_{x,y}}{4\pi}\frac{1}{d^2}
\end{equation}
with
\begin{equation}
  K = \frac{\mu_0(IL)}{2\pi(B\rho)},
\end{equation}
where $d$ is the distance between the wire and the beam, $\mu_0$ the permeability of the
vacuum, $(IL)$ the integrated wire strength, and $(B\rho)$ the beam rigidity.

Note that we take a positive sign for $d$ for a wire above the beam, and a
negative sign below the beam. We also assume that the reference vertical
orbit position at the location of the wire is zero ($y_{ref}=0$)
for the wire current off.
The sign of $K$ depends on the direction of the wire current
relative to the beam direction, and the charge of the beam particles.
In our case the wire current
has the opposite direction to the beam, the Blue wire is above and the Yellow
wire below the beam, and the beam particles have positive charges. In this
case the sign of $K$ is negative in Blue, and positive in Yellow. The orbit
change $\Delta y$ at the location of the wire due to the dipole kick
$\Delta y'$, for $\Delta y \ll d$, is then
\begin{equation}\label{eq:dist}
  \Delta y = \frac{K\beta_y}{2d} \frac{\cos{(\pi Q_y)}}{|\sin{(\pi Q_y)|}}.
\end{equation}
If the wire comes close to the beam Eq.~(\ref{eq:dist}) becomes inaccurate
and needs to be replaced by
\begin{equation}
  \Delta y = \frac{d}{2} -
  \sqrt{\frac{d^2}{4}-\frac{1}{2}K\beta_y\cot(\pi Q_y)}
\end{equation}
where $d$ is now the distance between the wire and the beam position at zero
wire current.

\begin{figure}[t]
\centering
\includegraphics[width=58mm,angle=-90]{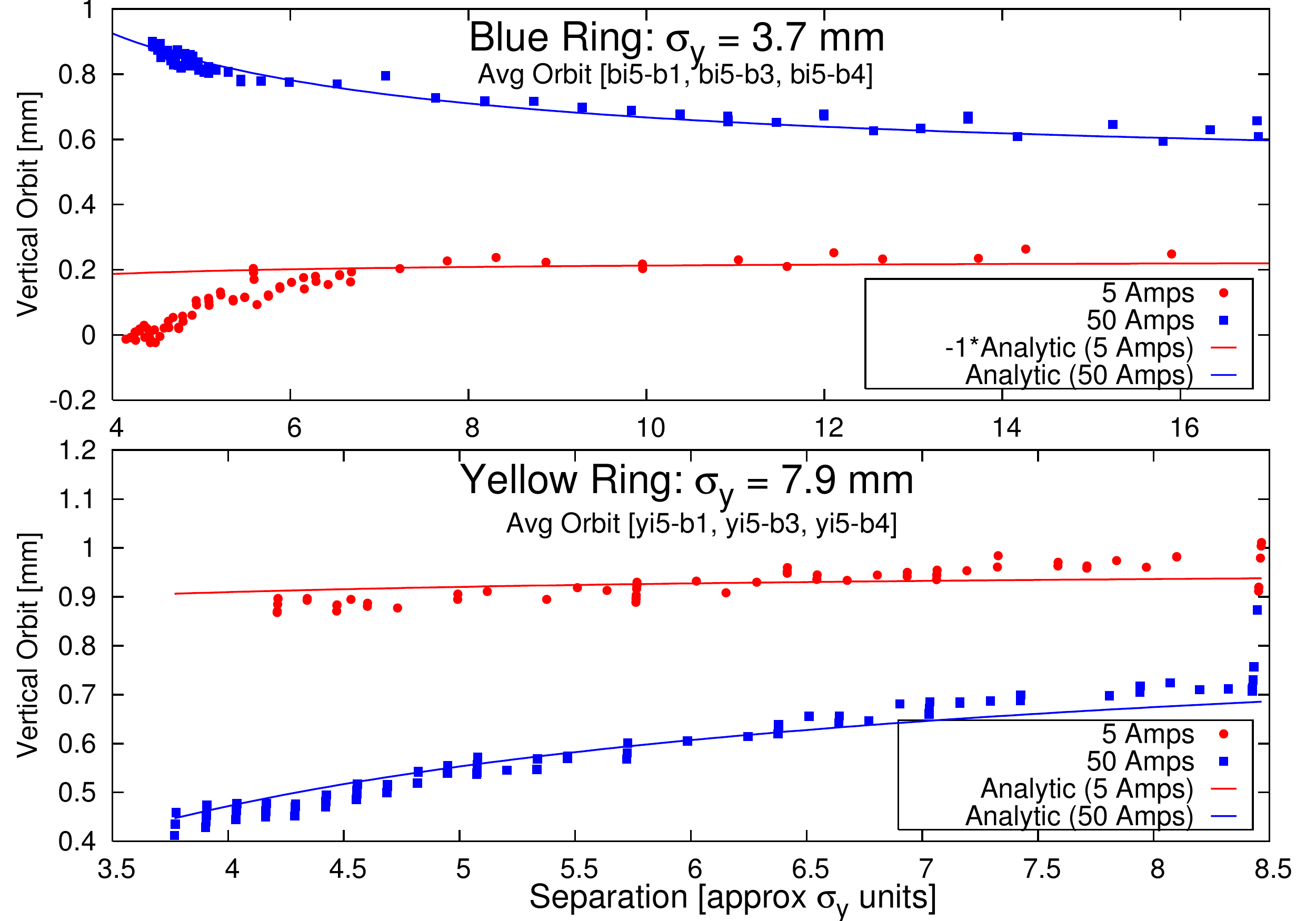}
\caption{Vertical orbit change (average of 3 BPMs near wire) as a function
of vertical distance, in Blue and Yellow ring at 5~A and 50~A.
Solid lines in all plots represent the analytical prediction.}
\label{fig:orbitvsd1}
\end{figure}

Orbit and tune changes agree with expectations under well controlled
experimental circumstances~\cite{Fisc2,Kim0}. Figure~\ref{fig:orbitvsd1} shows
a comparison of the measured beam trajectories to the analytical prediction as
a function of the separation between the wire and the Blue beam.

\begin{figure}[htb]
\centering
\includegraphics[width=58mm,angle=-90]{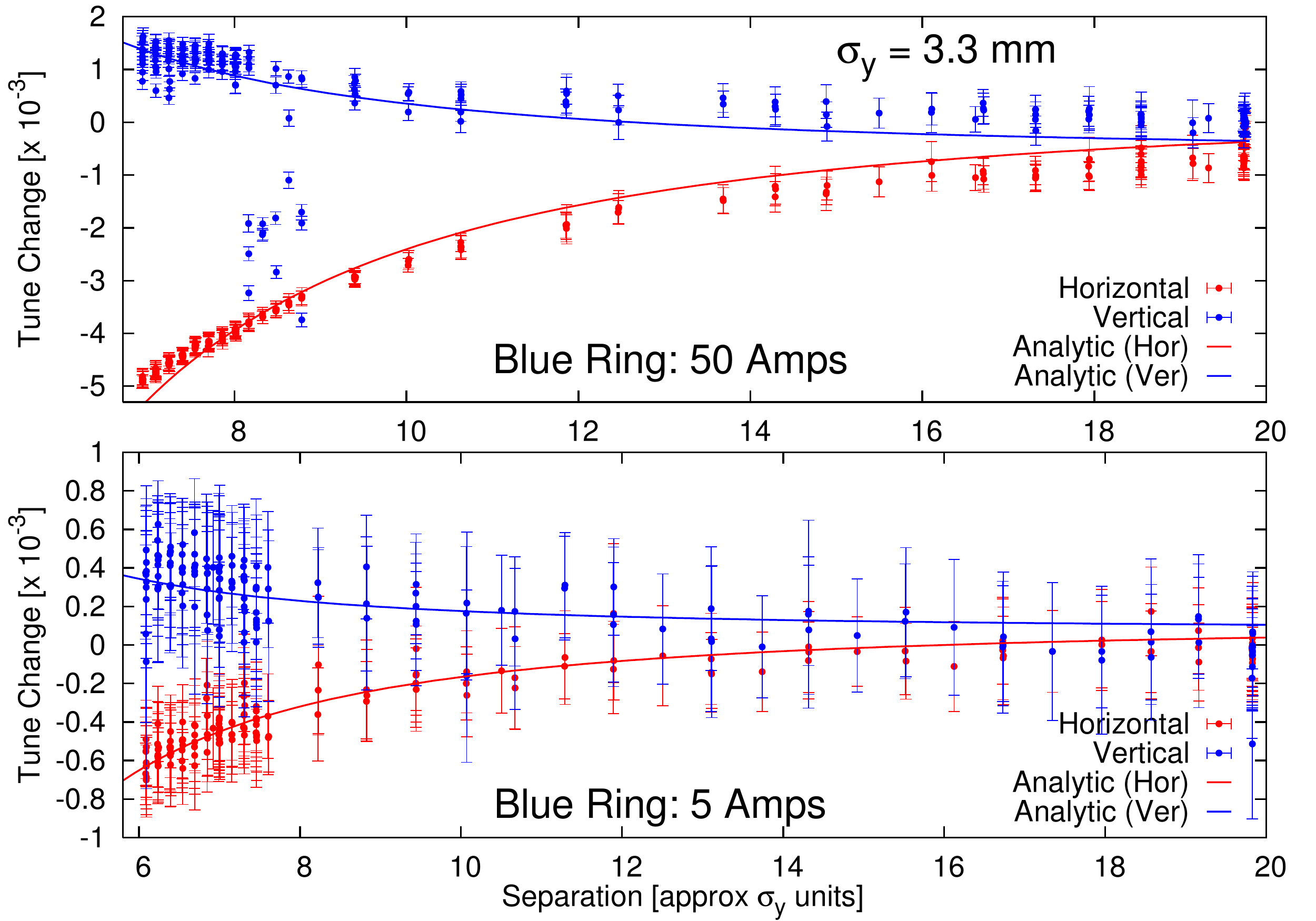}
\caption{Horizontal and vertical tune change for 50~A and 5~A wire current,
for the Blue ring.  Solid lines in all plots represent the analytical
prediction.}
\label{fig:tunevsd2}
\end{figure}

Figure~\ref{fig:tunevsd2} shows a comparison of the measured tunes to the
analytical prediction as a function of the separation between the wire and the
beam.

\begin{figure}[htb]\centering
\includegraphics[width=58mm,angle=-90]{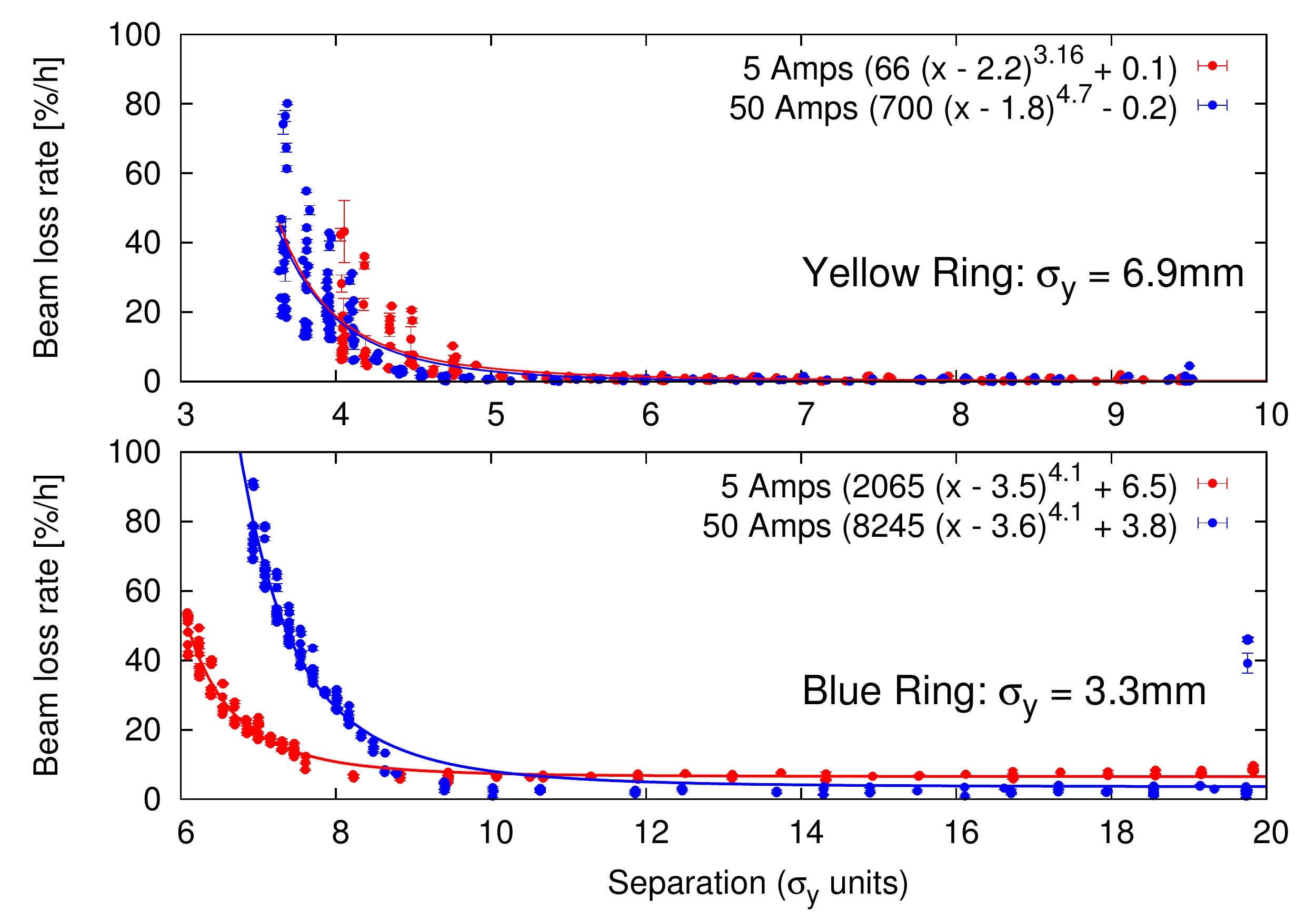}
\includegraphics[width=48mm,angle=-90]{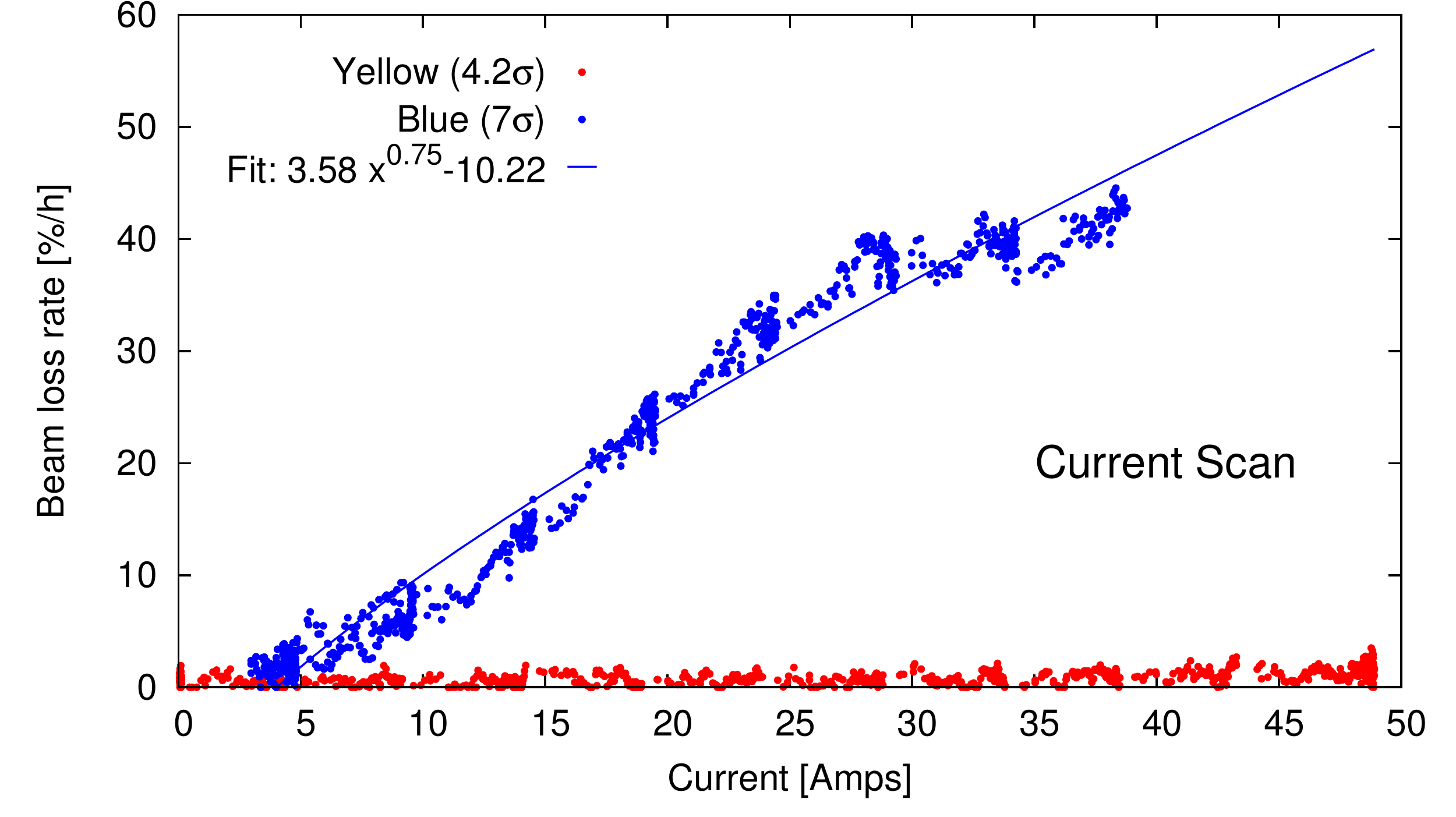}
\caption{Top: Beam loss as a function of the DC wire separation to the Blue and
the Yellow beams at 5~A and 50~A. Bottom: Beam loss due to a current scan in
the DC wire fixed at a given distance from the beam. Solid lines in all plots
show a power law fit to the losses.}
\label{fig:DISSCAN}
\end{figure}

The beam lifetime, however, is determined through the nonlinear beam--beam
effect and can only be assessed in detailed simulations.
Figure~\ref{fig:DISSCAN} (top) shows the beam loss rate as a function of the
vertical wire distance to the beam. The onset of losses due to a long-range
type interaction between the wire and the beam is visible. Similarly the effect
on beam losses due to a current scan at a fixed distance is shown in
Fig.~\ref{fig:DISSCAN} (bottom). The approximate separation in the Blue
ring is 9~$\sigma$ and in the Yellow ring is 5~$\sigma$.
The Yellow ring shows very weak or no effect
with a current scan which is probably due to a previous distance
scan resulting in a cleaning of the large amplitude particles.

\begin{figure}[htb]\vspace{-1.8cm}
\centering
\includegraphics[width=58mm,angle=-90]{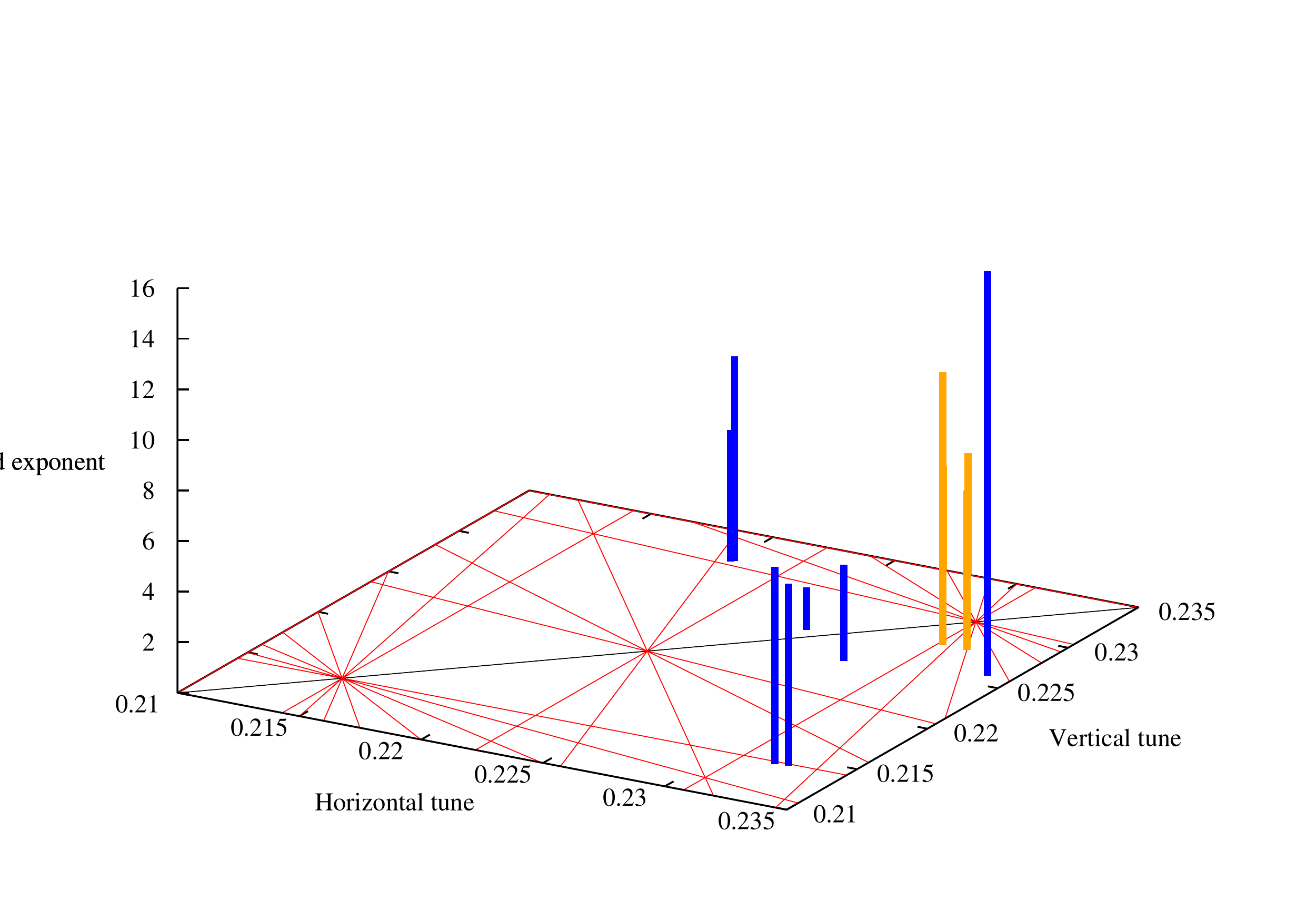}\vspace{-2cm}
\includegraphics[width=56mm,angle=-90]{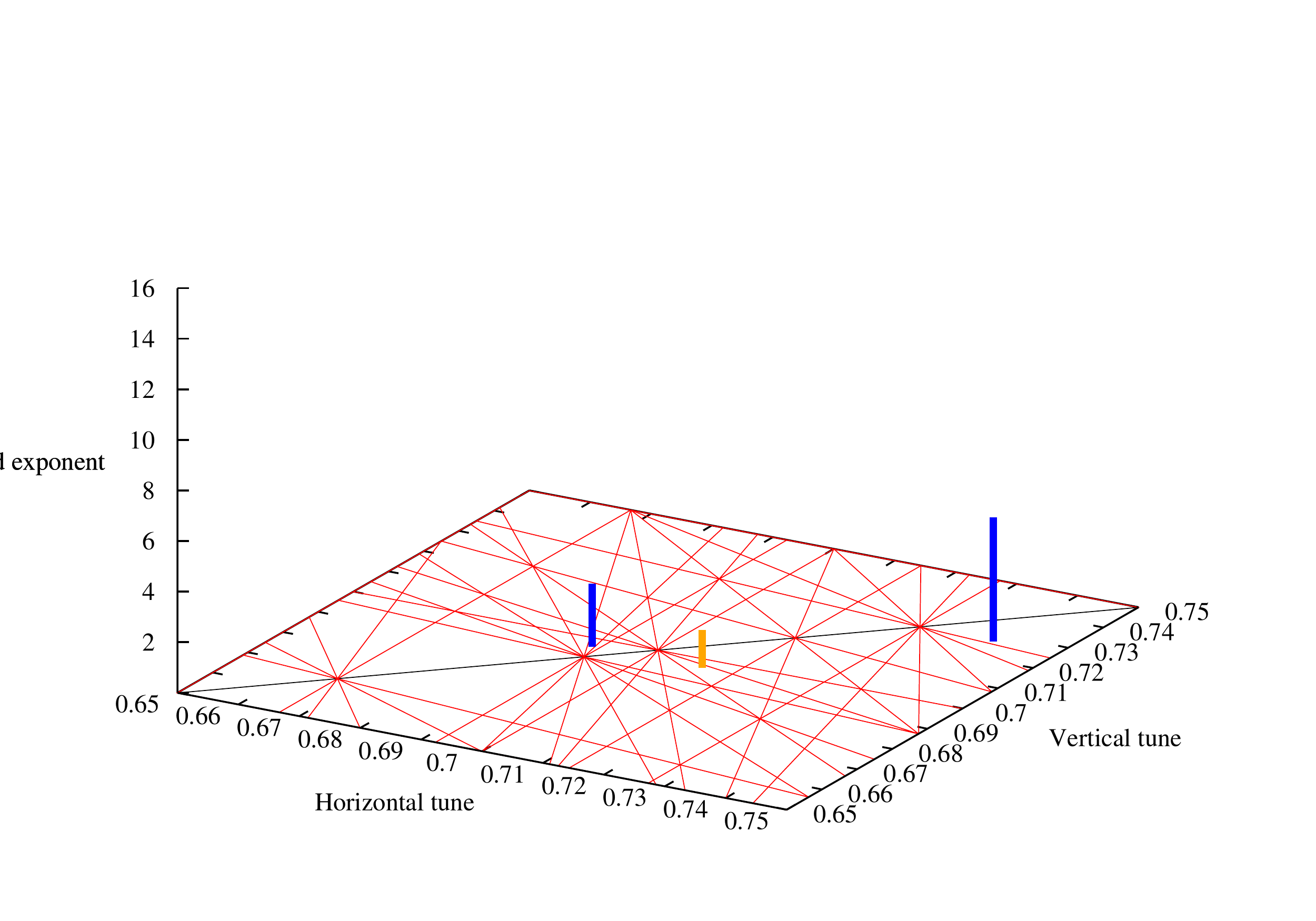}
\caption{Fitted exponents $p$ for long-range beam--beam experiments as a
function of the ion tunes (top) and the proton tunes (bottom). The fitted
exponents range from 1.7 to 16.}
\label{fig:alpha_vs_Q}
\end{figure}

It was speculated that the beam lifetime $\tau$ can be expressed as
$\tau = Ad^p$ where $A$ is an amplitude, $d$ the distance between wire and
beam, and $p$ an exponent that would typically be in a narrow range. For
the SPS $p$ had been found to be about 5, and for the Tevatron to be
about 3~\cite{Zimm8}. In Table~\ref{tab:LR} the fitted exponents are listed
for all cases for which a fit was possible. The fitted exponents range from
1.7 to 16, i.e. $p$ is not constrained within a narrow range. Ten of the thirteen
$p$ values are between 4 and 10. Figure~\ref{fig:alpha_vs_Q} shows the fitted
exponents $p$ as a function of the ion tunes in the upper part, and the proton
tunes in the lower part. Ion tunes near the diagonal and away from either
horizontal or vertical resonances show smaller exponents $p$. The experiments
also showed that the beam lifetime is reduced with increased
chromaticity~\cite{Fisc2}.

Another simple measure of assessing the long-range beam--beam effect in
experiments is the distance between the beam and wire (or other beam) at which
the beam lifetime becomes smaller than a certain value. We have chosen this
value to be 20~h, which would imply a luminosity lifetime of 10~h or less.
Table~\ref{tab:LR} shows an amplitude range between 3.5 and 17~$\sigma$. With
the available amount of data no clear correlation can be established between
this distance and the fitted coefficient $p$. In two cases the distance was found
to be as large as or larger than 10~$\sigma$, and most cases fall between 4 and
10~$\sigma$. Operation with less than 5~$\sigma$ separation appears to be
difficult~\cite{Abre1}. Note that the beam is sometimes used for multiple scans
and that a large lifetime drop at large distances is more typical for
previously unused beams (Table~\ref{tab:LR}).

\begin{figure}[htb]
\centering
\includegraphics[width=88mm,height=65mm]{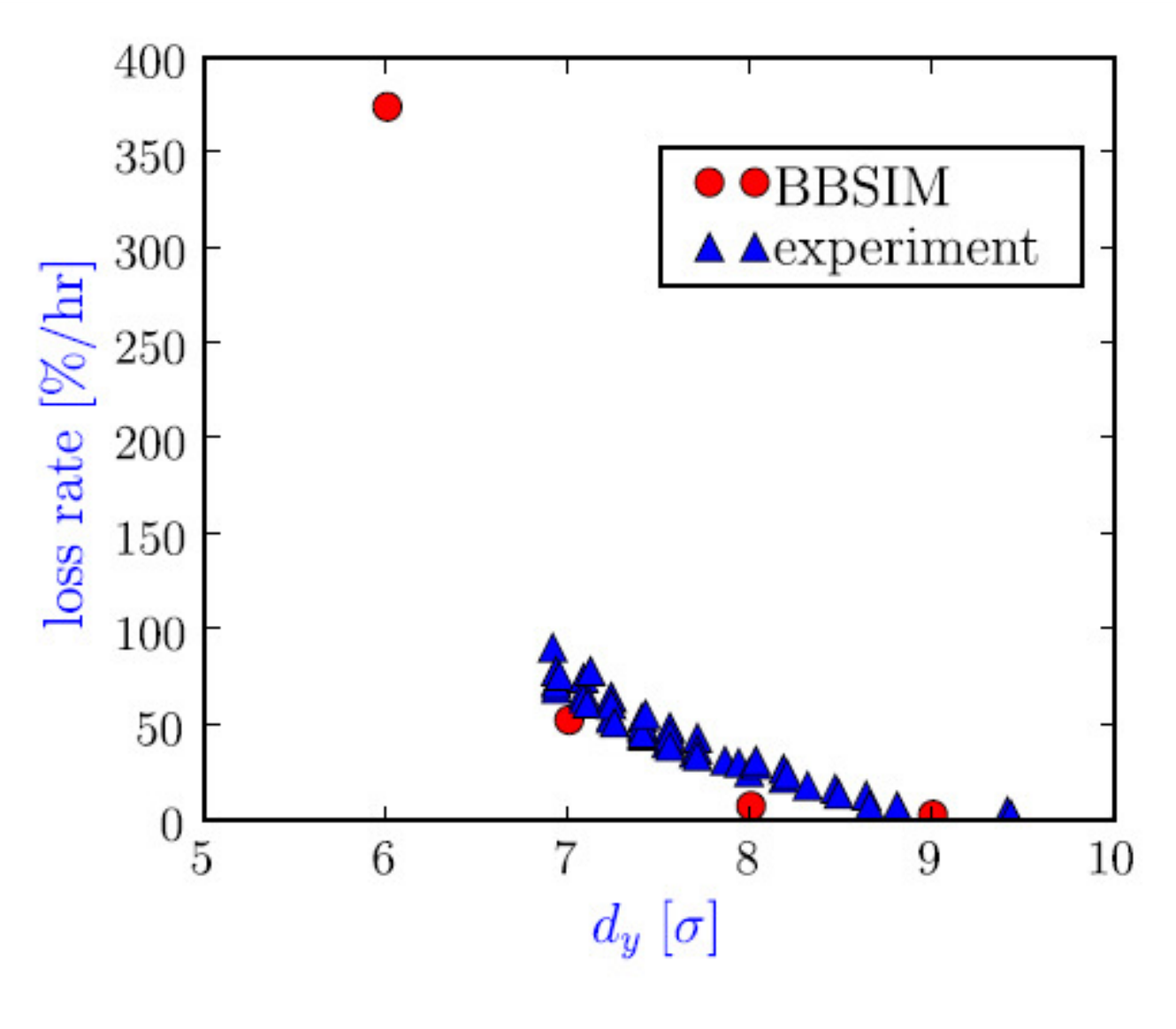}
\caption{Comparison of measured and simulated beam loss rate as a function
of distance between wire and beam. Experiment with gold beam at store,
wire strength of 125~Am~\cite{Kim2,Kim4}.}
\label{fig:bblr_sim}
\end{figure}

One important goal of the experiments is to benchmark simulations. In
several simulations the onset of large losses as a function of the distance
between wire and beam was reproduced within about 1~$\sigma$~\cite{Kim0,
Kim1,Kim2,Kim4,Kim3,Kabe,Dord4}.
One such comparison is shown in Fig.~\ref{fig:bblr_sim}.

\subsection{Long-range Effects with Head-on Collisions}
End of physics fills were initially used to test the effect of the wires on
colliding gold and deuteron beams (see Table~\ref{tab:LR}).
It should be noted that the beam--beam parameter of proton beams in RHIC
is approximately three times larger than the beam--beam parameter
of heavy ion beams.
The first dedicated experiment with protons to compare
the effect of the wire on colliding beams and compensation of a single
LR beam--beam interaction was conducted in 2009 at 100 GeV.
Due to aperture considerations for decreasing $\beta^*$, the Blue wire was
removed during the shutdown after the 2009 run and the Yellow wire was
removed subsequently. Therefore, the experiments in 2009
serve as the final set of measurements for LR beam--beam with
RHIC as a test bed. The relevant RHIC beam and lattice parameters
are listed in Table~\ref{tab:2009} for the experiments in 2009.
\begin{table}[htb]
\centering
\caption{Relevant RHIC beam and lattice parameters for experiments
with proton beams.}
\label{tab:2009}
\begin{tabular}{|l|c|c|c|}
\hline
Quantity  & \hspace*{-2mm}Unit & \hspace*{-2mm}Blue &
\hspace*{-2mm}Yellow \\ \hline\hline
Beam energy $E$                & GeV  & \multicolumn{2}{c|}{100}  \\
Rigidity $(B\rho)$             & Tm     & \multicolumn{2}{c|}{333.5}\\
Number of bunches              & -    & \multicolumn{2}{c|}{36}   \\
\# of colliding bunches        & -  & \multicolumn{2}{c|}{30}   \\ \hline
%parameter $K$ (at 50~A)        & -     & \multicolumn{2}{c|}{$-30.1$} \\
Bunch intensity      & $10^{11}$ & 1.7 & 1.7 \\
%beam loss rate w/o wire    & \%/h   & 2.5  & 1.5  \\
Norm. Emittance $\epsilon_{x,y}$   & $\mu$rad & 25,24 &  49,19 \\
%Ver rms beam size at wire     & mm     & 3.3  & 6.9  \\
Horizontal tune $Q_x$          & ...    & 28.691 & 28.232 \\
Vertical tune $Q_y$            & ...    & 29.688 & 29.692 \\\cline{3-4}
Chromaticities ($\xi_x,\xi_y$) & ...    & \multicolumn{2}{c|}{$(+2,+2)$} \\
\cline{3-4}
%Harmonic number $h$            & ...    & \multicolumn{2}{c|}{360} \\
%Gap voltage $V_{gap}$          & MV     & \multicolumn{2}{c|}{0.3} \\
%$\beta_x$ at wire location     & m      & 556  & 1566 \\
$\beta_x$ at wire location     & m      & 1566  & 556 \\
%$\beta_y$ at wire location     & m      & 1607 & 576 \\ \hline
$\beta_y$ at wire location     & m      & 576 & 1607 \\ \hline
\end{tabular}
\end{table}

Prior to a long-range compensation attempt, a position scan
of the wire on each beam was performed with a wire current of 50 A.
A 36$\times$36 bunch pattern with six non-colliding bunches was chosen
to enable a comparison of the lifetime in the presence of the
wire between single beam and colliding beams simultaneously.
%Figure~\ref{HONOHOBTF} shows the vertical beam transfer functions
%before and after head-on collisions were established.
The corresponding beam loss rates as a function of beam to wire separation
on both colliding and non-colliding bunches were measured. The initial
beam loss rates with colliding beams were stabilized to the
nominal 10\% per hour. The maximum total beam losses for the wire movements
towards the beam at fixed current were constrained to 100--150\% per hour 
for a very short period to avoid disrupting the beam quality
significantly for subsequent measurements.
%\begin{figure}[htb]
%\centering
%\includegraphics[width=57mm,angle=-90]{HONHO_BTF-eps-converted-to.pdf}
%\caption{Beam transfer function measurements in the vertical plane
%for Blue (top) and Yellow (bottom) rings with.
%Comparison between bunches with head-on and no head-on collision is shown.}
%\label{HONOHOBTF}
%\end{figure}

\begin{figure}[htb]
\centering
\includegraphics[width=45mm,angle=-90]{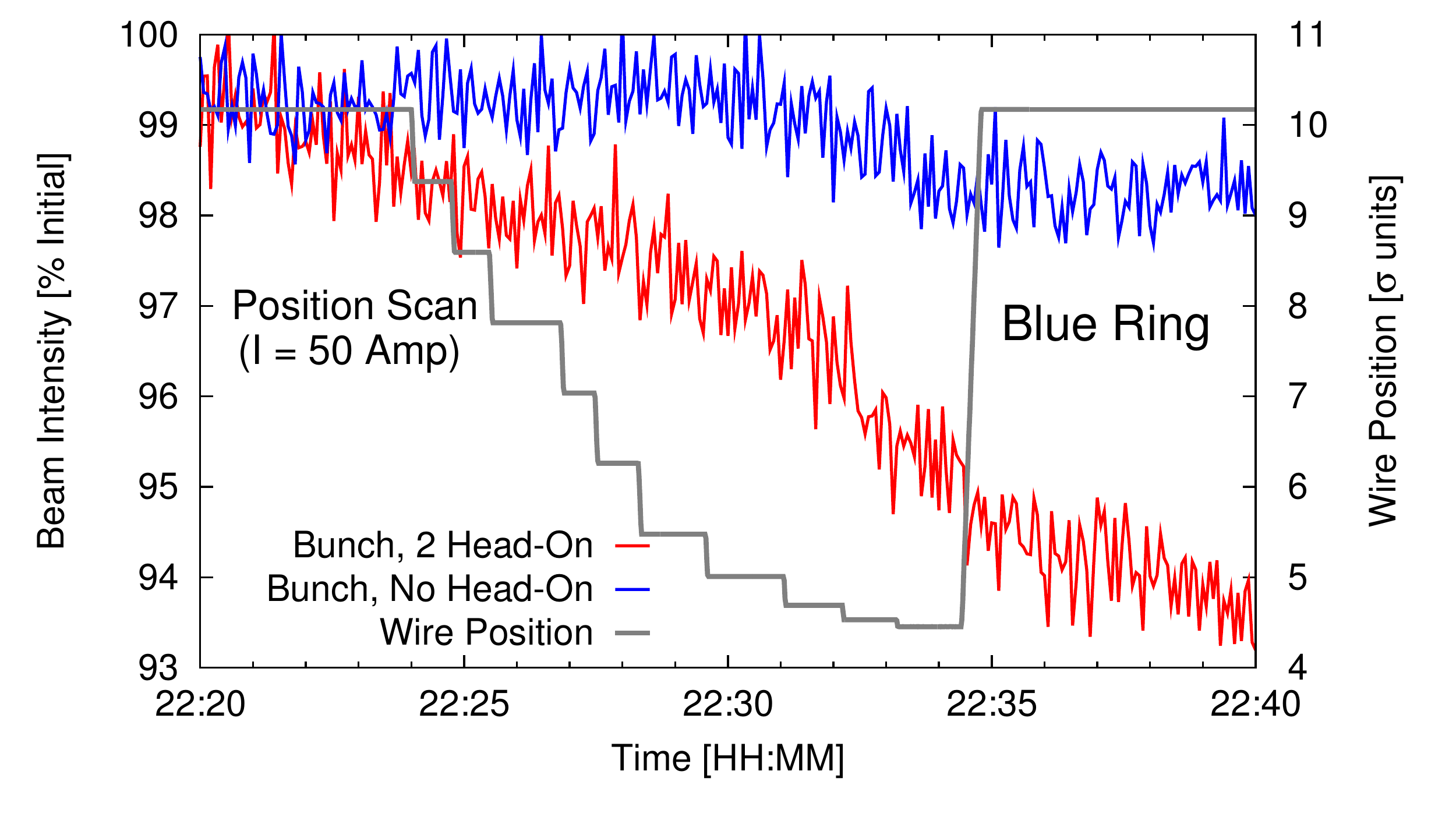}
\includegraphics[width=45mm,angle=-90]{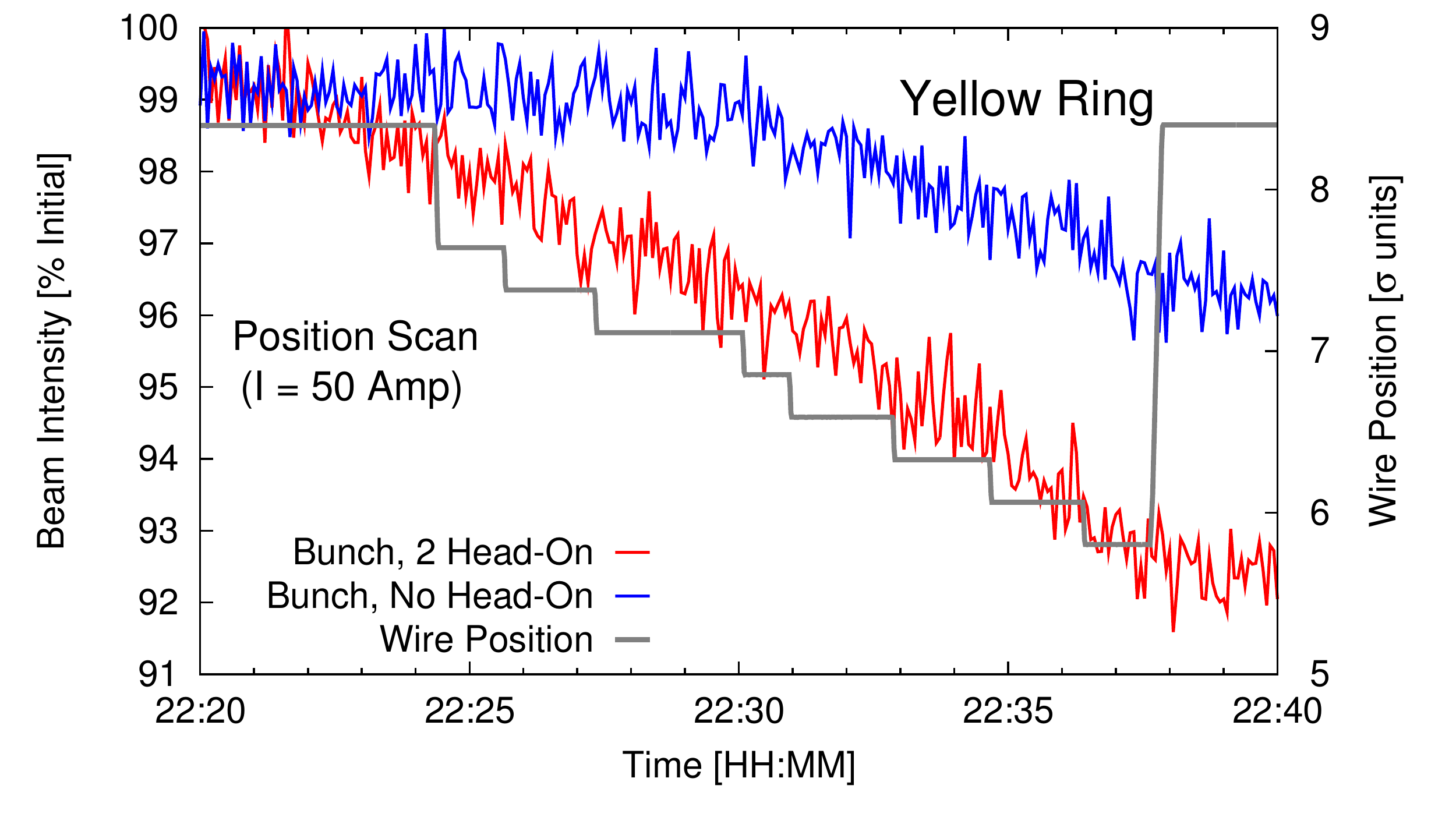}
\caption{Single bunch intensities as a function of wire position
for Blue (top) and Yellow (bottom) rings with a current of 50 A.
Comparison between bunches with head-on and no head-on collision is shown.}
\label{HONOHO}
\end{figure}

Figure~\ref{HONOHO} shows the evolution of the intensity
between bunches with and without head-on collisions.
It is evident that the bunches with the head-on collisions have
a more severe effect from the LR forces of the wire.
Several hypotheses can be formulated to explain the increased losses
for bunches with head-on collision.
\begin{itemize}
\item The dynamic aperture for the bunches with head-on is significantly
smaller than that of the single beam which could lead to the
observed beam losses.
\item It was also suggested by an anonymous referee 
that the addition of the head-on collisions enhances the diffusion
leading to enhanced losses in the presence of long-range interactions.
Figure~\ref{HONOHO} clearly shows a larger initial slope
for bunch intensities with collisions. However, it is difficult to
untangle the contribution from the reduced dynamic aperture 
as opposed to enhanced diffusion.
\item The additional tune shift due to the wire along with
large head-on tune shift could lead to beam losses due to very
limited tune space available. No tune optimization was performed
during the experiment.
\item The effect of the wire on the orbit
can introduce a static offset between the two beams at the IP
which is approximately proportional to the wire distance.
A large offset due to the kick
from the wire can lead to emittance blow-up and beam losses~\cite{Piel}.
The relative offset at the collision point during the wire scan
with 50 A (see Fig.~\ref{ORBITOFF}) is well below the 1~$\sigma$ level
which is very small.
\begin{figure}[htb]
\centering
\includegraphics[width=46mm,angle=-90]{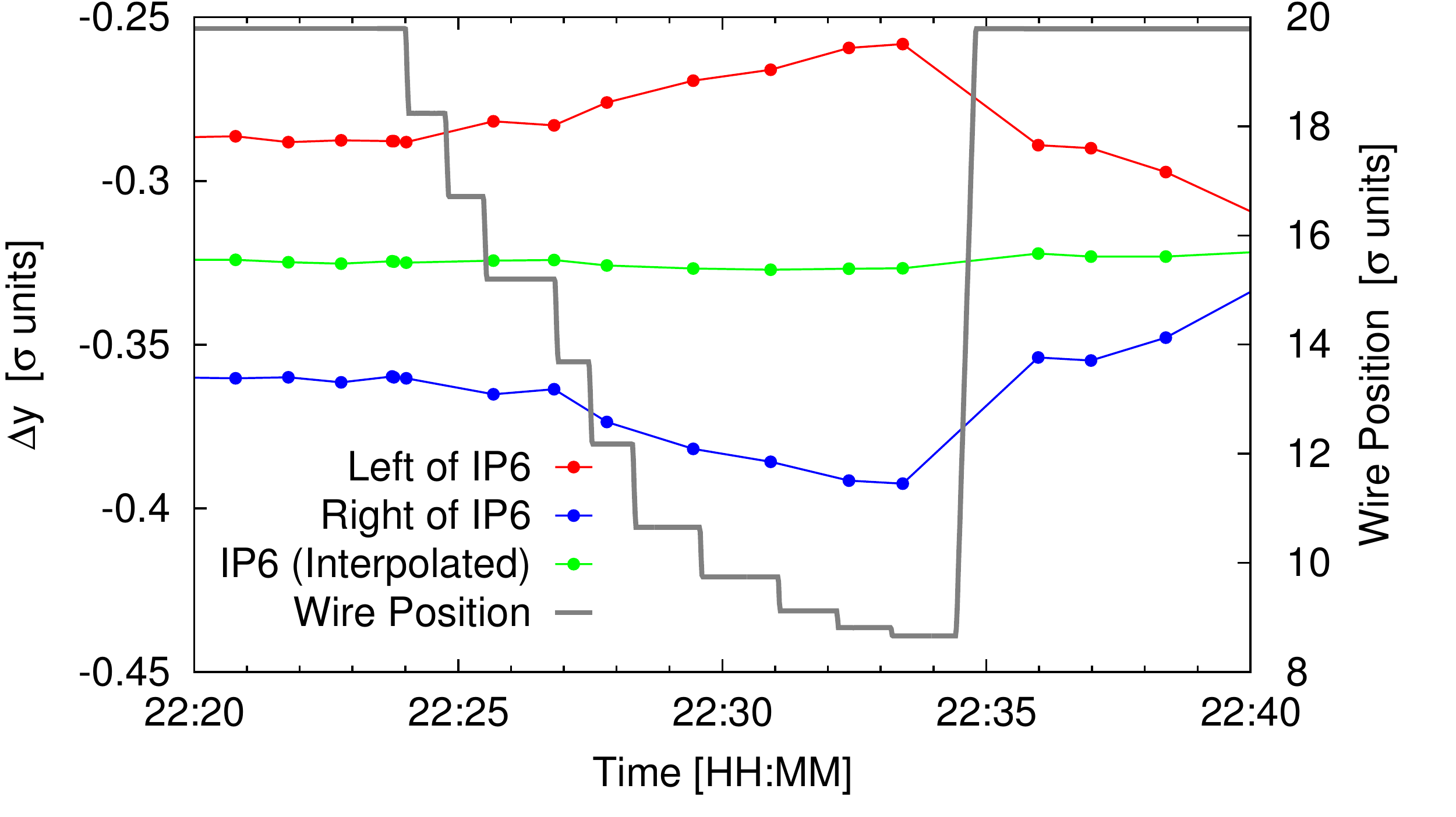}
\includegraphics[width=46mm,angle=-90]{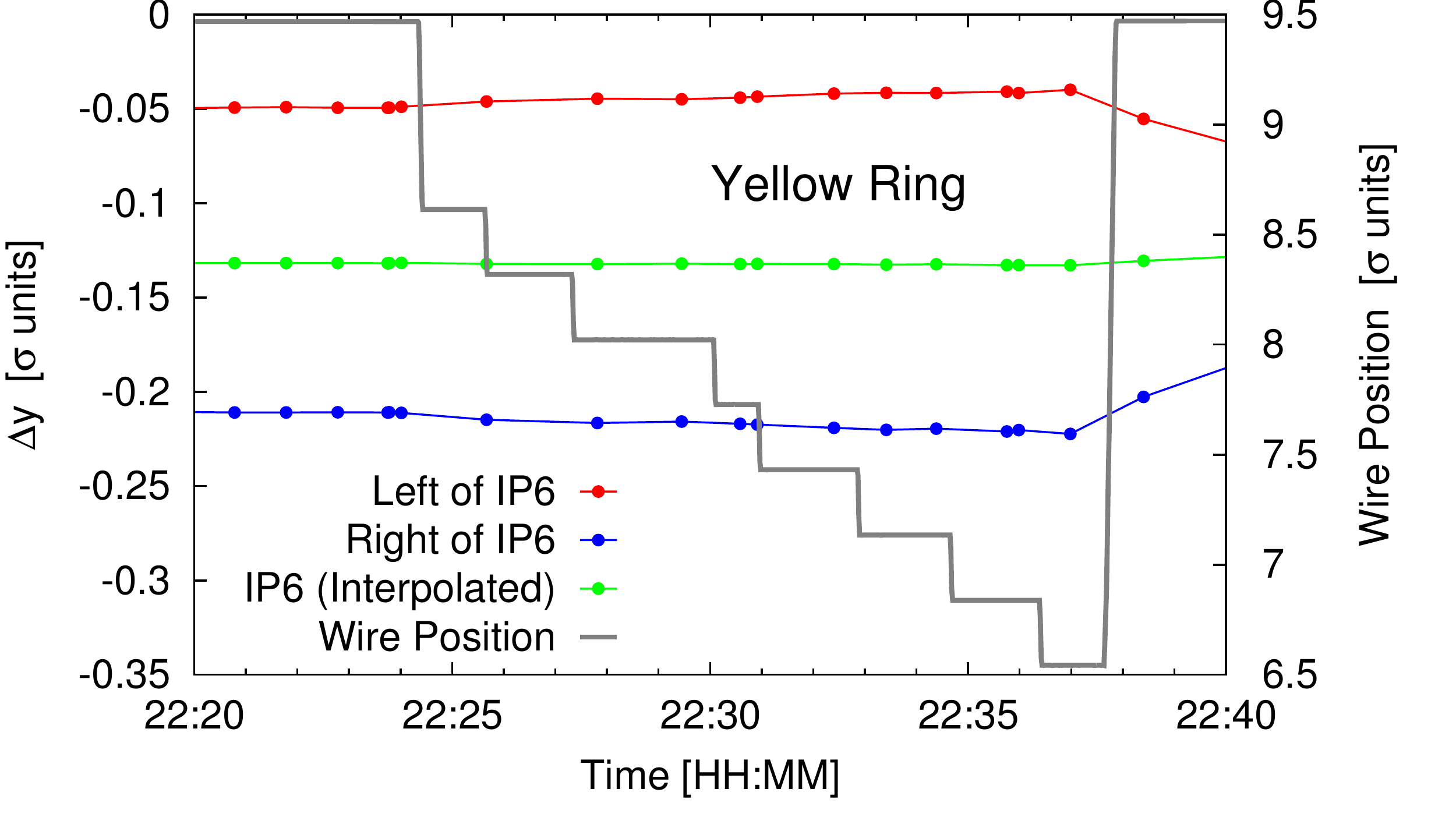}
\caption{Orbit offset at IP6 as a function of the wire position
for Blue (top) and Yellow (bottom) rings with a current of 50 A.}
\label{ORBITOFF}
\end{figure}
\end{itemize}
However, simulations to support each of the above hypotheses to explain
its contribution towards observed losses is beyond the scope of this paper.

\subsection{Single Long-range and Wire Compensation}
\begin{figure}[htb]
\centering
\includegraphics[width=45mm,angle=-90]{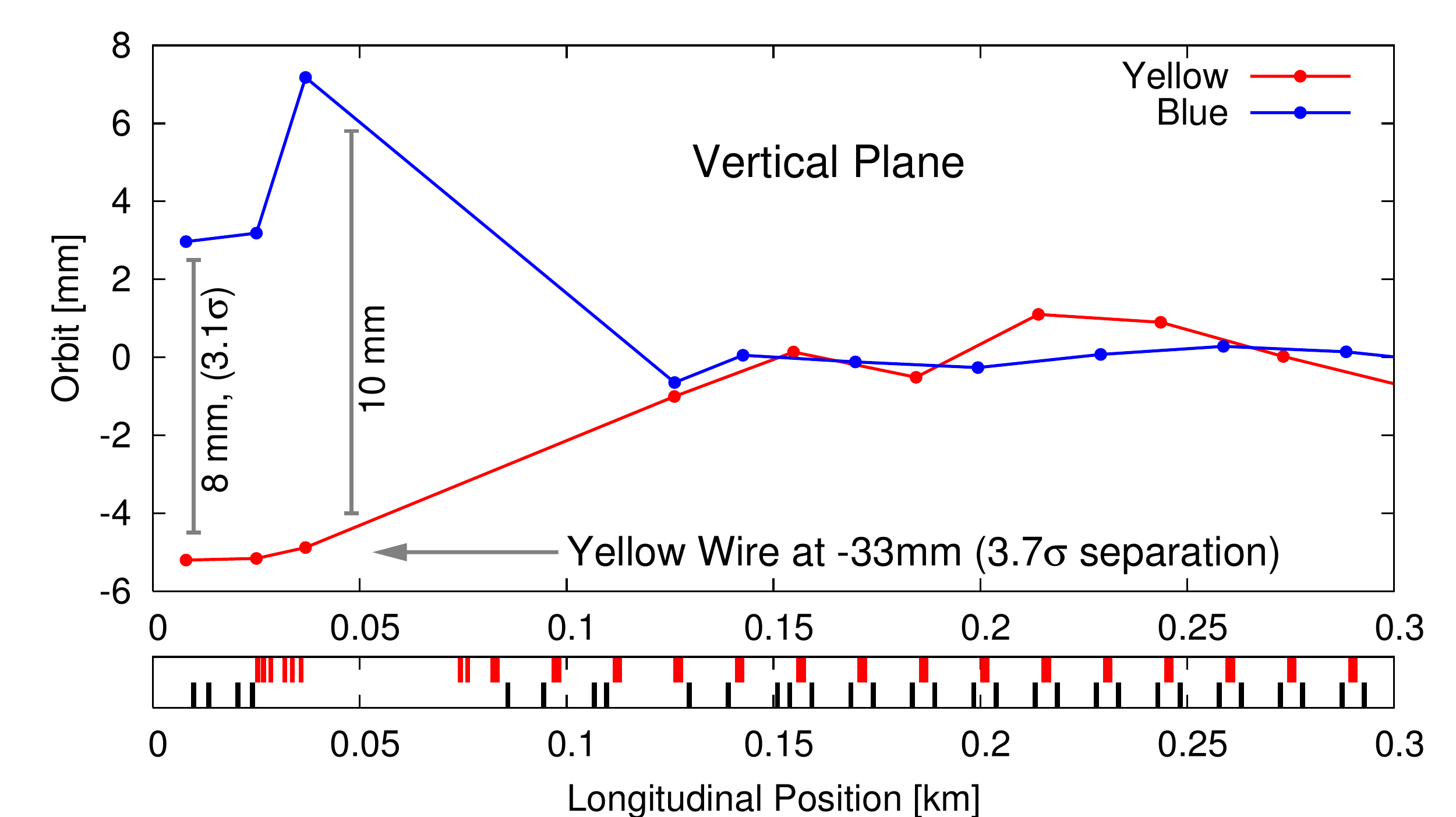}
\caption{Orbits right of IP6 for the Blue and the Yellow ring
with the LR interaction setup near the DX magnet at approximately
3.1~$\sigma$.}
\label{fig:ORBIT}
\end{figure}

The bunch spacing and the interaction region geometry in RHIC
does not inherently have LR beam--beam interactions. It is therefore
necessary to shift the collision point towards the DX magnet closest
to the DC wires as noted before. This location enables an
artificially induced LR interaction between the two beams and
simultaneously allows for a minimum phase advance between the LR
interaction and DC wires (6 deg). Additionally, this location has
sufficient aperture for an orbit scan with the range of interest
(3--10~$\sigma$). Figure~\ref{fig:ORBIT} shows the trajectories of the Blue
and Yellow rings with the LR interaction set at approximately 3.1~$\sigma$.

\begin{figure}[htb]
\centering
\includegraphics[width=45mm,angle=-90]{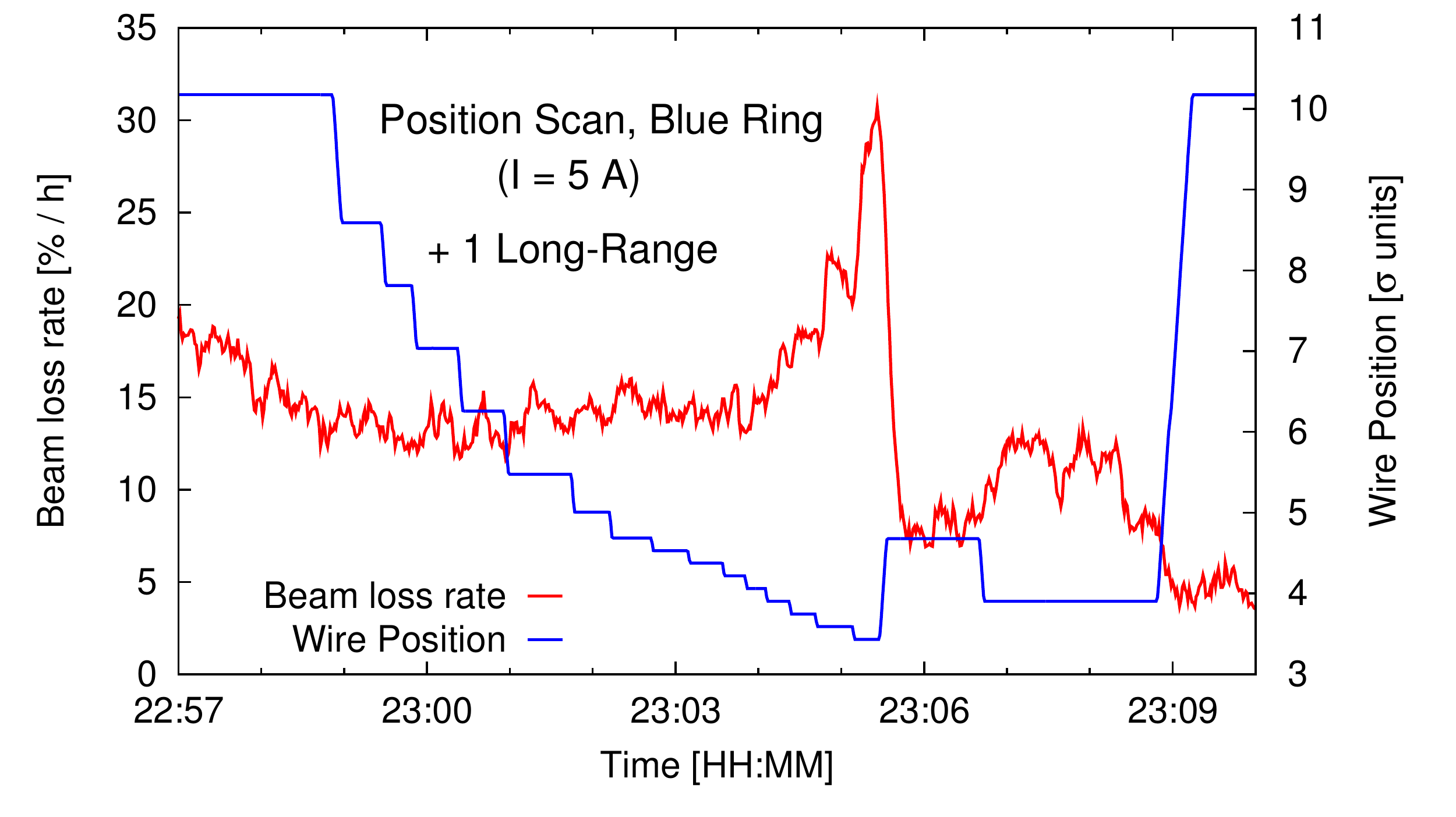}
\includegraphics[width=45mm,angle=-90]{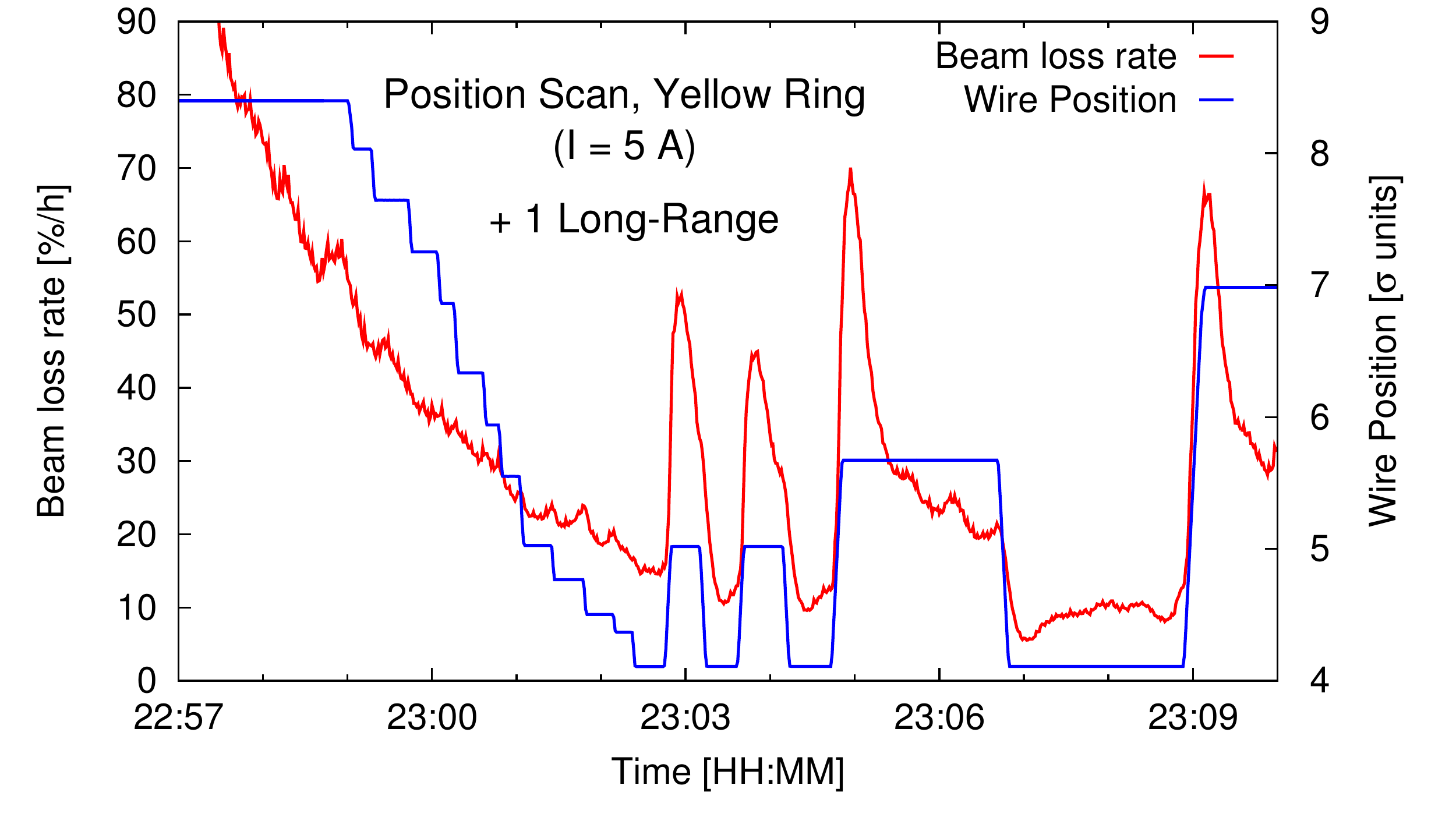}
\caption{Beam loss rate for a Blue (top) and Yellow (bottom) bunch with one
long-range interaction, and an additional wire interaction. The wire position
varies, and the wire current is constant at 5~A.}
\label{LONOLO}
\end{figure}

The individual bunch intensities and beam losses were recorded during the
position scan with the LR compensation~\cite{Cala}. Figure~\ref{LONOLO} shows
the beam losses as a function of the wire position. In the Blue ring, the
losses are always increasing as the wire approaches the beam.
Therefore, no
evidence of compensation of the LR interaction from the Blue beam is visible.
However, in the Yellow ring, the beam lifetime improved as the beam to wire
distance approaches 3~$\sigma$ (Fig.~\ref{LONOLO}). Consecutive retractions
and restoration of the beam to wire distance to 3~$\sigma$ show similar
improvement of the beam lifetime. This indicates a compensation of the effect
of LR interaction by the DC wire.

\begin{figure}[htb]
\centering
\includegraphics[width=48mm,angle=-90]{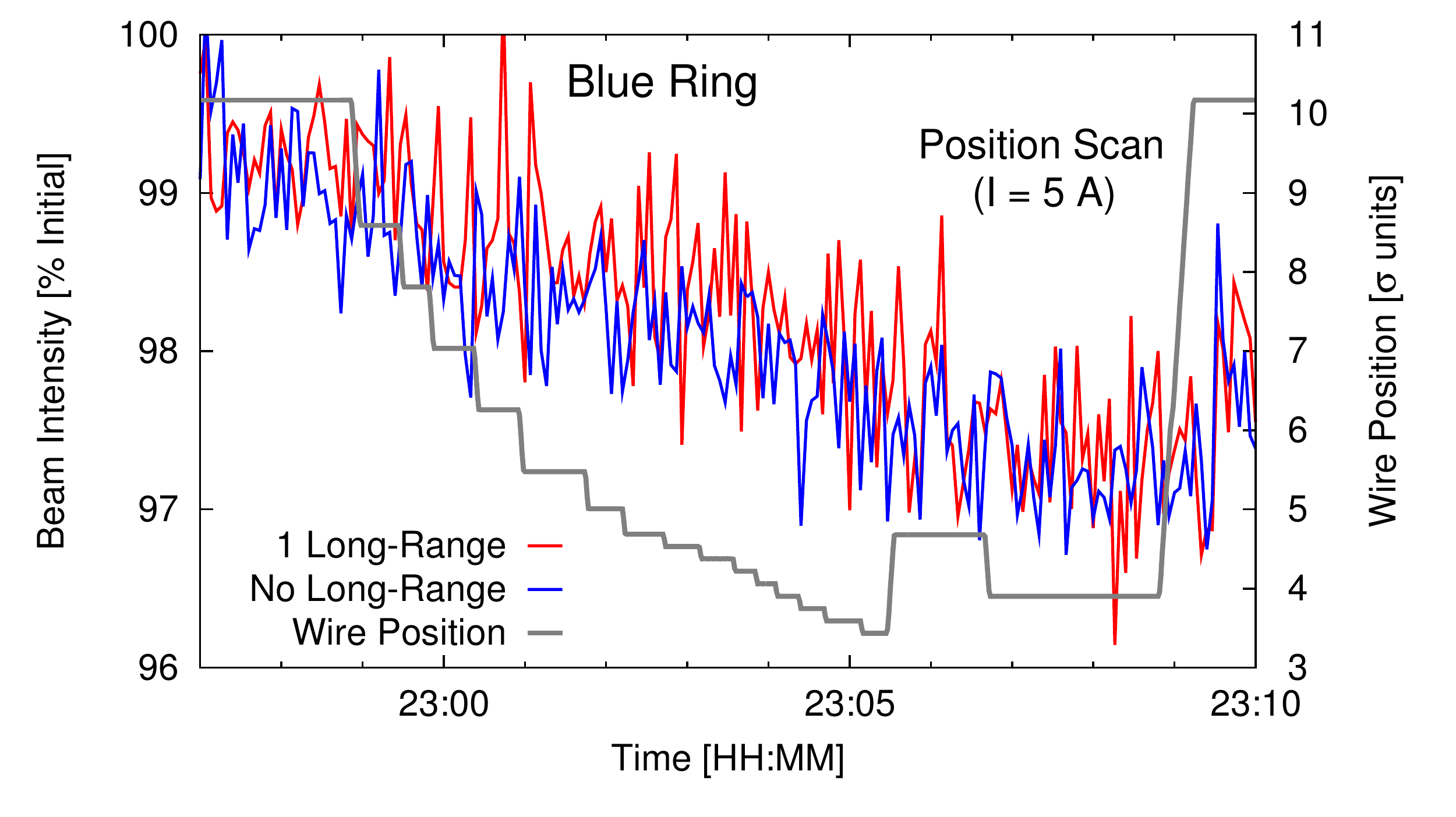}
\includegraphics[width=48mm,angle=-90]{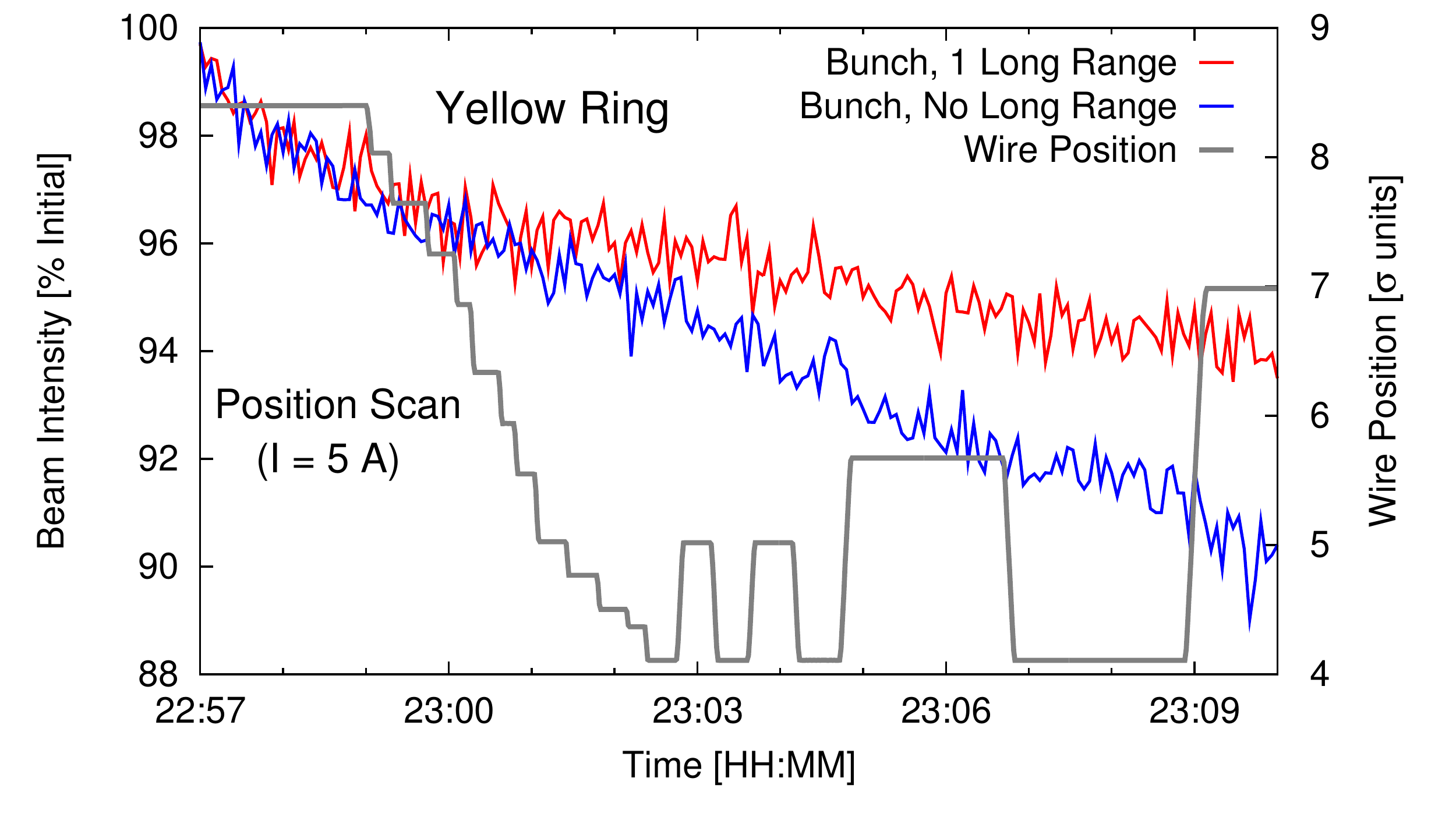}
\caption{Beam intensity comparison between bunches with a single long-range
and no long-range interaction as a function of the wire position for Blue
(top) and Yellow (bottom) rings with a wire current of 5 A.}
\label{COMP}
\end{figure}

In addition to beam losses, the individual bunch intensities with
and without LR interactions and simultaneous compensation is shown
in Fig.~\ref{COMP}. Note that all 36 bunches experience the effect
of the DC wire, but only 30 bunches experience LR interactions.
Therefore, only bunches with an LR interaction can experience a
compensation. In the Blue ring, the bunch intensity evolution is
similar for bunches with and without LR compensation. Hence, only
the effect from the wire is visible. The bunches with LR interaction
and simultaneous compensation have reduced beam losses as compared
to the bunches that only see the wire. This is consistent with the
beam loss measurements (Fig.~\ref{LONOLO}).

\section{Summary}
Long-range beam--beam experiments were conducted in RHIC from 2005 to 2009.
The motivation for these were two-fold. First, the experimental data can
benchmark simulation codes for situations of strong localized long-range
beam--beam interactions as they will exist in the LHC. Second, the compensation
of a single long-range beam--beam interaction can be tested in a scheme that
is also usable in the LHC.

These experiments complement the experience with long-range beam--beam
interactions in the Sp$\mathrm{\bar{p}}$S and Tevatron, wire experiments in
the SPS, and the partial long-range compensation in DA$\Phi$NE. The RHIC wires
created strong localized long-range beam--beam effects, comparable in strength
to the effect expected in the LHC, with a beam that has a lifetime typical of
hadron colliders, and including head-on beam--beam collisions.
The observed orbit and tune changes due to the wire were as expected.
The effect of the long-range beam--beam interactions on the beam loss rate
is sensitive to a number of beam parameters such as the tunes and
chromaticities. Fitting the beam lifetime $\tau$ to an exponential function
$\tau \propto d^p$ as a function of the distance $d$ between the beam and the
wire, exponents $p$ in the range between 1.7 and 16 were found. Distances
smaller than 5~$\sigma$ created losses too large for collider operation.
The experimentally observed distance from the wire to the beam at which large
beam losses set in could be reproduced in simulations within 1~$\sigma$.
The beam lifetime with long-range interactions created by the wire
was degraded further through head-on collisions.
%showed clear evidence of degradation in the presence
%as opposed to without head-on collisions.
A single attempt to compensate long-range beam--beam interaction via
a DC wire showed evidence of compensation.

\section{Acknowledgements}
For discussions and help we are most thankful to collaborators within the US
LHC Accelerator Research Program as well as several people at various
laboratories. Among these are
O.~Br\"{u}ning, R.~DeMaria, U.~Dorda, W.~Herr, A.~Kabel, H.-J.~Kim,
J.-P.~Koutchouk, C.~Milardi, K.~Ohmi, T.~Pieloni, J.~Qiang, F.~Schmidt,
T.~Sen, G.~Sterbini, and F. Zimmermann.

Work supported by Brookhaven Science Associates, LLC under Contract
No. DE-AC02-98CH10886 with the US Department of Energy, and
in part by the US LHC Accelerator Research Program.

\end{document}